\documentclass[amsmath,amssymb,aps,pra,reprint,superscriptaddress,footnoteinbib]{revtex4-2} 
\usepackage{amsmath,amssymb}
\usepackage[english]{babel}
\usepackage{ bbold }
\usepackage{upgreek}
\usepackage{dsfont}
\usepackage{graphicx}
\usepackage{comment}
\usepackage{braket}
\usepackage{colortbl}
\usepackage{ragged2e}
\usepackage{blindtext}
\usepackage{hyperref}
\begin{document}

\title{Quantum rotations of nanoparticles}

\author{Benjamin A. Stickler}
\affiliation{University of Duisburg-Essen, Faculty of Physics, Lotharstra\ss e 1, 47048 Duisburg, Germany}
\affiliation{Imperial College London, Quantum Optics and Laser Science, Exhibition Road, London SW7 2AZ, United Kingdom}
\email{benjamin.stickler@uni-due.de}

\author{Klaus Hornberger}
\affiliation{University of Duisburg-Essen, Faculty of Physics, Lotharstra\ss e 1, 47048 Duisburg, Germany}

\author{M. S. Kim}
\affiliation{Imperial College London, Quantum Optics and Laser Science, Exhibition Road, London SW7 2AZ, United Kingdom}

\begin{abstract}
Rotations of microscale rigid bodies exhibit pronounced quantum phenomena that do not exist for their center-of-mass motion. By levitating nanoparticles in ultra-high vacuum, researchers are developing a promising platform for observing and exploiting these quantum effects in an unexplored mass and size regime. Recent experimental and theoretical breakthroughs demonstrate exquisite control of nanoscale rotations, setting the stage for the first table-top tests of rotational superpositions and for the next generation of ultra-precise torque sensors. Here, we review the experimental state of the art and discuss promising routes towards quantum rotations.
\end{abstract}

\maketitle

\thispagestyle{empty}

\tableofcontents

\section{Introduction}
{Spinning tops are fascinating toys, amusing children and grown-ups alike.} Ultimately, this is due to the non-linear nature of rigid-body revolutions \cite{goldstein}. It gives rise to textbook phenomena such as gyroscopic stabilization, precession, and the mid-axis instability. In the quantum realm, rotors show pronounced interference effects in their orientation, which have no analogue in the centre-of-mass motion of the object and no correspondence  in its classical rotations. This renders orientational quantum interference an attractive path towards  testing and exploiting quantum physics with massive objects.

Researchers in the field of levitated optomechanics \cite{millen2020} are perfecting their techniques: State-of-the-art experiments control nanoparticle rotations with ultra-high precision \cite{hoang2016,kuhn2017a,kuhn2017b,rashid2018}, spin them with unprecedented rotation frequencies \cite{reimann2018,ahn2018,jin2020}, achieve record-breaking torque sensitivities \cite{ahn2020}, and cool nanoparticles into their center-of-mass ground state \cite{delic2020,magrini2020,tebbenjohanns2021quantum}. In a typical setup, a sub-micron dielectric particle is optically, electrically, or magnetically levitated in ultra-high vacuum and coherently interfaced with a laser field, serving to cool and manipulate the nanoparticle motion. First experimental demonstrations of rotational cooling \cite{delord2020,bang2020,vanderlaan2020b} bring tests of microscale quantum rotations within reach.

This Perspective reviews the experimental state of the art by explaining how the rotational dynamics of trapped particles can be controlled and exploited. We present promising routes towards demonstrations of quantum rotational effects, including orientational quantum revivals \cite{stickler2018b}, quantum persistent tennis-racket flips \cite{ma2020}, and spin-stabilized magnetic levitation \cite{rusconi2017a}. We conclude by discussing how quantum rotations can be exploited for the next generation of macroscopic quantum superposition tests \cite{schrinski2019} and high-precision torque sensors.

\section{Recent progress on controlling rotations in high vacuum}

Current experiments align nanoscale rotors with millirad precision  \cite{kuhn2017a,kuhn2017b,ahn2018,reimann2018,rashid2018}, spin them at GHz frequencies \cite{reimann2018,ahn2018,jin2020}, and cool their rotations to sub-Kelvin temperatures \cite{delord2020,bang2020,vanderlaan2020b} {above the ground state}. Controlling and cooling microscale rotations requires novel strategies to handle their peculiar motion, as described by the non-linear Euler equations for the angular velocity in presence of non-harmonic torques, see Box 1. 
This trait of rigid-body revolutions can also be exploited technologically. For instance, the gyroscopic stabilization of rapidly spinning nanorotors allows one to frequency-lock their mechanical motion to an external drive, enabling pressure sensing with unprecedented precision \cite{kuhn2017b}. Other manifestations of rotational non-linearities already observed with levitated particles include precession \cite{rashid2018,bang2020}.

Approaches towards controlling and cooling mechanical rotations can be grouped according to how they couple to the control fields, i.e. whether the particles are dielectric, magnetizable, electrically charged or permanent magnets -- see Fig.\,\ref{fig:Fig1}. In the following, we first discuss strategies for aligning and spinning levitated particles, before presenting techniques to cool their mechanical rotation into the quantum regime, as required for sensing applications and fundamental tests.

\subsection{Aligning and spinning microscale particles}

\vspace{2mm}

\noindent {\bf Optical control of dielectric rotors---} A popular platform for levitating neutral dielectric objects are linearly polarized optical tweezers \cite{gieseler2020}. Since the laser wavelength typically exceeds the particle extensions, they tend to align their axis of maximal susceptibility with the local field polarization, see Fig.\,\ref{fig:Fig1}(a). The optical anisotropy  can be either material-induced, caused by optical birefringence, or shape-induced, due to deviations from perfect sphericity. Such anisotropies are unavoidable given that perfectly spherical and optically isotropic objects are impossible to fabricate. For markedly aspherical particles the maximal susceptibility can appreciably exceed that of an isotropic body. Aligning such particles with an external field can thus significantly enhance their centre-of-mass trapping potential and force sensitivity \cite{kuhn2015,stickler2016a}.

Trapped rotations of aligned particles, dubbed {\it librations} {(see Box 1)}, have been observed with buffer-gas-cooled rotors  of various shapes \cite{hoang2016,kuhn2017a,rashid2018,vanderlaan2020}. The resulting librational dynamics can  deviate strongly from a two-dimensional harmonic motion \cite{rashid2018,bang2020} because rotations around the axis of maximal  susceptibility are not affected by the linearly polarized trapping field. As a consequence, the associated angular momentum is conserved, leading to rotational precession, see Box 2. This has been observed with symmetric rotors \cite{rashid2018,bang2020}, and was found to limit rotational feedback cooling schemes \cite{zhong2017,seberson2019}. Rotational precession can be controlled with elliptically polarized light fields \cite{schafer2020}, which exert a non-conservative radiation torque in addition to the conservative optical potential. {This torque is due to the scattering of light carrying angular momentum, and thus the rotational analogue of radiation pressure.}

\begin{figure*}
\hrule
\vspace{2mm}
\section*{{Box 1: Orientation, alignment, rotation, and libration}}
\begin{minipage}{0.75\textwidth}
\justify
{The {\bf orientation} of a rigid rotor may be specified by three orientational degrees of freedom describing how a space-fixed frame of reference is rotated into a body-fixed one. A popular choice to specify the rotation matrix are Euler angles in the $z$-$y'$-$z''$ convention. The natural body-fixed frame is provided by the rotor's {\bf principal axes}, which are the eigenvectors of the {\bf inertia tensor}, reflecting its mass distribution. The rotor is said to be {\bf aligned} with a space-fixed direction whenever the latter lies parallel or anti-parallel to a body-fixed axis. Aligning two principal axes simultaneously fixes the third one and thus orients the body up to an inversion.}
\end{minipage}
\begin{minipage}{0.24\textwidth}
\begin{flushright}
\includegraphics[width = 0.85\textwidth]{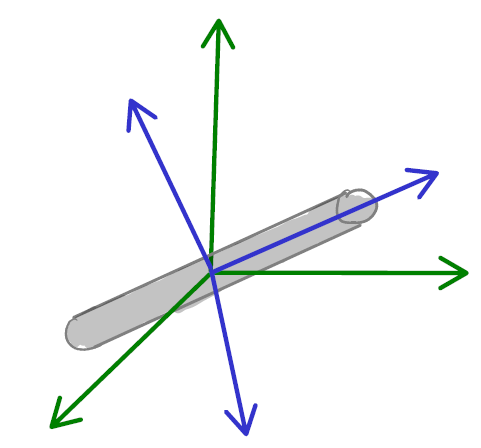}
\end{flushright}
\end{minipage}

\vspace{2mm}

\noindent
\begin{minipage}{0.24\textwidth}
\begin{flushleft}
\includegraphics[width = 0.85\textwidth]{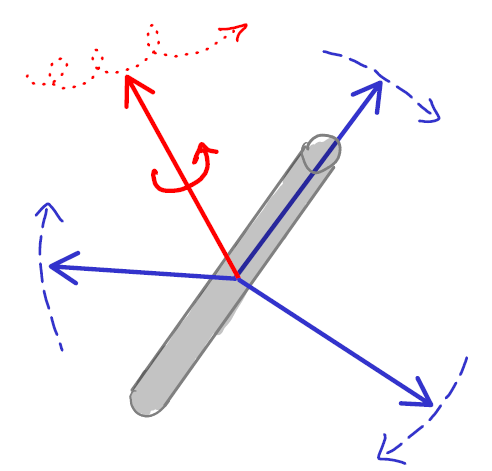}
\end{flushleft}
\end{minipage}
\begin{minipage}{0.75\textwidth}
\justify
{The classical {\bf rotation dynamics} of the body-fixed frame are described by the {\bf angular velocity} vector, whose three body-fixed components obey the coupled non-linear {\bf Euler equations}. This non-linearity of rotations ensures that the {\bf angular momentum} vector, the product of inertia tensor and angular velocity vector, is conserved in the absence of external torques. If a torque keeps one body-fixed axis strongly aligned to a space-fixed direction, the resulting rotation dynamics are dubbed {\bf librations}. In this case, two angles are trapped harmonically, while Euler's equations can still couple them non-linearly to the angular velocity for rotations around the body-fixed frame.}
\end{minipage}

\vspace{2mm}

\begin{minipage}{0.75\textwidth}
\justify
{The {\bf quantum state} of a rigid rotor can be put into a superposition of different orientations. Due to the bounded configuration space the canonically conjugate momenta take discrete values, the uncertainty product of orientation and angular momentum depends on the quantum state of the rotor, and the components of the angular momentum operator do not commute (but form a Lie algebra). As a consequence, the energies of a free rigid top always depend non-linearly on the quantum numbers, giving rise to rotational interference phenomena (see Box 4).} 
\end{minipage}
\begin{minipage}{0.24\textwidth}
\begin{flushright}
\includegraphics[width = 0.85\textwidth]{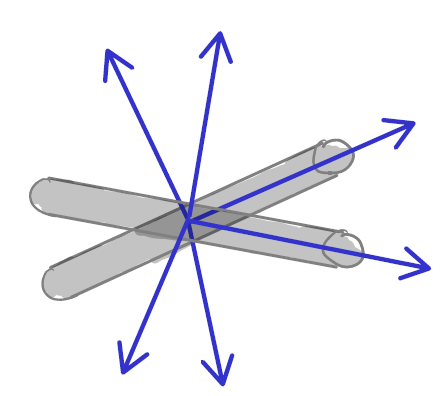}
\end{flushright}
\end{minipage}
\hrule
\end{figure*}

The effect of the radiation torque is best illustrated for circular laser polarization. In this case, the optical potential confines the rotor to the polarization plane \cite{kuhn2017a}, while light scattering induces an angular acceleration within this plane, see Fig.\,\ref{fig:Fig1}(b). Fast rotations enhance the transverse laser confinement through the gyroscope effect (Box 2). The maximal rotation speed is determined by gas damping \cite{arita2013,kuhn2017a,kuhn2017b,monteiro2018,martinetz2018} and ultimately by material stresses \cite{hummer2020}. Experimentally demonstrated rotation rates reach GHz-frequencies \cite{ahn2018,reimann2018,ahn2020,jin2020},  where centrifugal forces start deforming the particle. 

Elliptically polarised laser fields yield both a conservative optical potential, aligning the rotor within the polarization plane, and a non-conservative torque,  aiming to propel the particle within this plane. Importantly, the conservative potential is characterized by both polarization directions, thus breaking the axial symmetry of the linear field polarization, see Fig.\,\ref{fig:Fig1}(c). Tuning the degree of laser ellipticity such that the conservative torque dominates enables full three-dimensional alignment of aspherical bodies. This can be combined with a high-finesse optical cavity to realise cooling the particle to its combined rotational-translational quantum ground state \cite{schafer2020}.

Beyond three-dimensional alignment, the interplay of conservative and non-conservative optical torques has been put forward for torque sensing \cite{kuhn2017b,arita2020}. For instance, by periodically switching between circular and linear laser polarization the rotation frequency of a probe particle can be locked to the ticking of an external clock. The relative phase between external drive and observed particle orientation depends sensitively  on the gas-induced frictional torque, which allows measuring the gas pressure with sub-percent accuracy \cite{kuhn2017b}.

\noindent {\bf Electric manipulation of charged rotors---} As a viable alternative to optical tweezers, Paul traps can be used to suspend charged nanoparticles \cite{kane2010,millen2015,delord2017b,nagornykh2017,bykov2019}{, whose charge state can be controlled \cite{moore2014,frimmer2017}.} The rotation of charged nanoparticles can be controlled with electrostatic fields or weak  lasers \cite{delord2020}. Adding superconducting circuitry might even allow to coherently network  levitated particles in all-electrical setups  \cite{goldwater2019,martinetz2020}.

The simplest Paul trap geometry consists of  an AC-driven ring centred at the trap axis and two grounded endcap electrodes. This gives rise to an oscillating electric quadrupole field, which generates an orientation-dependent potential determined only by the particle charge, dipole, and quadrupole moment \cite{delord2017a,martinetz2020}. The resulting trapping torque tends to align polar particles with the local field direction, thus coupling their rotational motion to the centre-of-mass dynamics \cite{martinetz2020}, see Fig.\,\ref{fig:Fig1}(d). The trap also exerts a torque due to the quadrupole moment, which is position-independent and which dominates for highly charged particles with negligible dipole moment. For such particles, the rotation dynamics decouple from the centre-of-mass motion, as in optical tweezers. Three-dimensional alignment  can be achieved with asymmetric quadrupole fields provided that the nanoparticle has three distinct quadrupole moments, see Fig.\,\ref{fig:Fig1}(e).

\begin{figure*}
\hrule
\vspace{2mm}
\section*{Box 2: The non-linearity of classical rotations} 
\begin{minipage}{0.8\textwidth}
\justify
The {\bf gyroscopic effect} describes that rapidly spinning bodies tend to retain their axis of rotation, which coincides with the angular momentum direction. Gyroscopic stabilization has many applications in engineering and navigation. For levitated particles, it allows separating the trapping frequencies of libration and center-of-mass motion and fixing the plane of rotation. Several experiments have observed and exploited this effect, for instance in the ultra-fast rotation of silicon nanorods \cite{kuhn2017a,kuhn2017b} and of nanodumbbells \cite{ahn2018}.
\end{minipage}
\begin{minipage}{0.19\textwidth}
\begin{flushright}
\includegraphics[width = 0.85\textwidth]{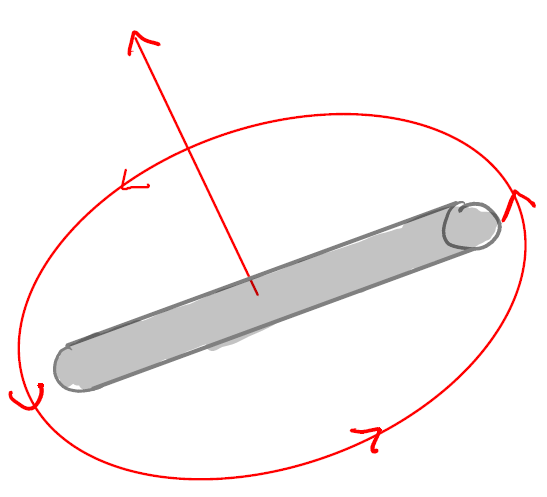}
\end{flushright}
\end{minipage}

\vspace{2mm}

\noindent
\begin{minipage}{0.19\textwidth}
\begin{flushleft}
\includegraphics[width = 0.85\textwidth]{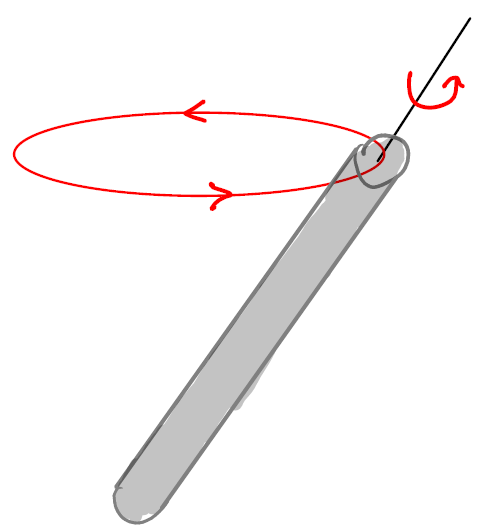}
\end{flushleft}
\end{minipage}
\begin{minipage}{0.8\textwidth}
\justify
A symmetric rotor shows {\bf precession} if its direction of angular velocity circles around a space-fixed axis. This motion can be  torque-free or torque-induced, with a speed proportional or  inversely proportional to the rotation rate around the symmetry axis, respectively. The latter type can be relevant for trapped particles if the trapping field tends to align the rotor symmetry axis with a space-fixed direction  \cite{rashid2018,bang2020}. Precession is in general superimposed with {\bf nutation}, a rocking motion of the angle to the space-fixed axis.
\end{minipage}

\vspace{2mm}

\noindent
\begin{minipage}{0.8\textwidth}
\justify
The {\bf mid-axis instability} describes the impossibility of freely rotating an asymmetric body around its axis of intermediate moment of inertia. Instead, the rotor reverses periodically its mid-axis orientation, referred to as the {\bf tennis-racket effect}. The frequency of these flips depends strongly on the initial state. The mid-axis instability will be important for interpreting experiments with freely revolving microscale particles, since perfect symmetry is impossible to achieve.
\end{minipage}
\begin{minipage}{0.19\textwidth}
\begin{flushright}
\includegraphics[width = 0.85\textwidth]{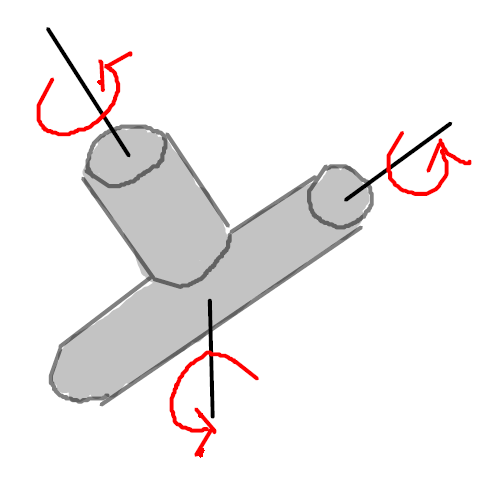}
\end{flushright}
\end{minipage}
\vspace{2mm}
\hrule
\end{figure*}

If the Paul trap is driven at high frequencies, the rotations of the trapped particle decompose into a slowly varying macrorotation and a small amplitude microrotation, wiggling with the drive frequency. Depending on the shape of the quadrupole field and the particle charge distribution, the time-averaged potential describing the macrorotation can have several local minima, offering great flexibility for designing the orientational potentials. For instance, in an azimuthally symmetric trap the particle main quadrupole axis can librate in direction of the trap axis or rotate freely around it, depending on the initial condition \cite{martinetz2020}. This way, planar rotations can be realized without the need of circularly polarized tweezers. 
 
Electric traps can be combined with optical setups and electric circuits. For instance, rotation frequency of nanoparticles can be locked to the Paul-trap drive by optically accelerating their rotation \cite{coppock2016}. When reaching a rotation rate sufficiently close to the AC drive the particles continue rotating with this frequency even after the laser is switched off. Furthermore, the rotation of polar particles towards an electrode induces an electric current in an attached circuit \cite{martinetz2020}, which can be fed into an electric feedback loop or dissipated in an RLC circuit, yielding all-electrical rotational cooling. An additional electrostatic field allows selecting the axis that couples to the circuit degrees of freedom. 

\vspace{2mm}

\noindent {\bf Controlling magnetic rotors---}  Magnetic traps provide an alternative to optical and electrical manipulation schemes requiring no driving fields \cite{millen2020}. There are several methods for trapping and aligning diamagnetic particles and permanent micromagnets. Like electric traps, magnetic techniques eliminate photon scattering  and open the door for networking levitated objects with superconducting circuits \cite{cirio2012,romeroisart2012,pino2018}. 
Moreover, magnetic fields affect the internal magnetization dynamics, which can strongly interact with mechanical rotations via the Einstein-de Haas \cite{einstein1915} and Barnett \cite{barnett1915} effects (Box 2).

Diamagnetic particles can be stably levitated in magnetic quadrupole fields \cite{hsu2016,slezak2018,obrian2019} with frequencies up to kHz. The induced magnetization is determined by the negative-definite magnetic susceptibility tensor, implying that the particle is attracted towards the trap centre, where the field vanishes. The associated magnetic torque tends to align the main susceptibility axis with the local field direction. However, shape and material-induced anisotropies can be small, rendering diamagnetic traps potentially useful for rotational quantum experiments. Diamagnetic levitation has been successfully demonstrated with nanodiamonds \cite{obrian2019} and silica nanoparticles \cite{slezak2018}, and has been proposed for superconducting spheres \cite{hofer2019,latorre2020chip}.

Permanent micromagnets can be levitated even without an external magnetic field. When placing a superconductor below a micromagnet, the former expels the magnetic field and thus exerts a repulsive force and a torque \cite{druge2014damping,timberlake2019,wang2019,vinante2020,gieseler2020b}. Cutting a non-circular hole into the surface can trap and align the magnet in-parallel to the superconducting surface \cite{pratcamps2017}, leaving it free to rotate around its magnetization axis. Trapping can also be achieved by cooling a type-II superconductor below its critical temperature while the micromagnet is present \cite{wang2019,gieseler2020b}. The levitated nanoparticle then interacts with a frozen dipole and simultaneously with its image dipole, yielding centre-of-mass and librational trapping \cite{kordyuk1998,wang2019}. A recent experiment \cite{gieseler2020b} in such a setup {studied} the influence of the  Einstein-de Haas effect on the librational trapping frequencies \cite{rusconi2016,rusconi2017a,rusconi2017b}{, and suggested that it can be observed in the near future}.

\begin{figure*}[t!]
\centering
    \includegraphics[width = 0.99\textwidth]{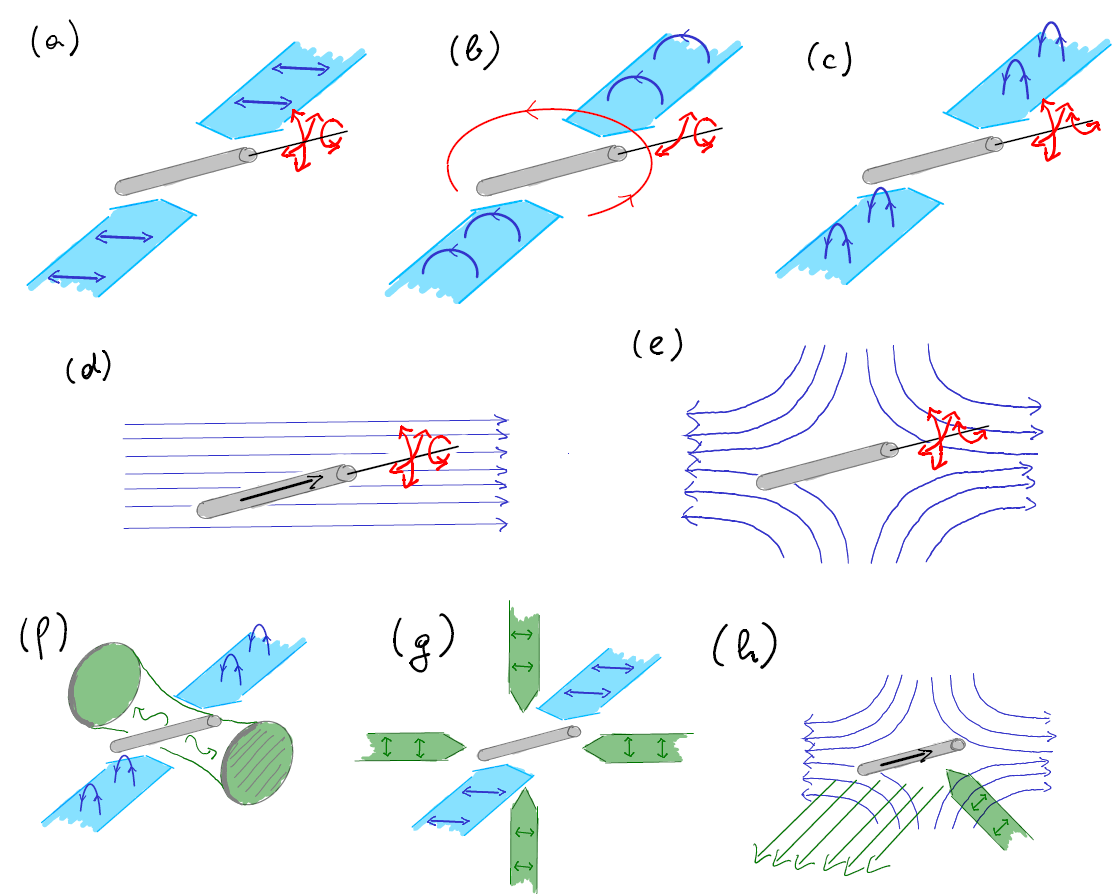}
    \caption{{\bf (a) Optical alignment:} Particles with an optical anisotropy tend to align their axis of maximum susceptibility with the linear tweezer polarization \cite{hoang2016,stickler2016a,kuhn2017a}, while rotations around it remain free. {\bf (b) Optical spinning:} If illuminated by circularly polarized light beams, the particle spins in the plane orthogonal to the beam axis \cite{kuhn2017a,kuhn2017b,ahn2018,reimann2018,jin2020,ahn2020}. Periodically switching between (a) and (b) can frequency-lock the nanoparticle rotations to the modulation frequency \cite{kuhn2017b}. {\bf (c) 3D optical alignment:} Full alignment of all body-fixed axes can be achieved by balancing the conservative and non-conservative torques exerted by elliptically polarised laser beams\cite{schafer2020}. {\bf (d) Aligning dipoles:} Permanent electric or magnetic dipoles can be aligned with homogeneous electrostatic or magnetostatic fields, which do not affect rotations around the dipole axis. {\bf (e) 3D Aligning quadrupoles:} Electric or magnetic quadrupole fields fully align permanent electric or magnetic quadrupoles along their main quadrupole axes \cite{delord2017a,martinetz2020}. {\bf (f) Elliptic coherent-scattering cooling:} Levitating an asymmetric rotor in an elliptically polarised tweezer, which couples coherently to two orthogonally polarised cavity modes. If the tweezer is red-detuned from the cavity resonance, the retarded back-action of the cavity  field on the particle motion can extract energy from all six degrees of freedom, enabling combined rotational-translational ground state cooling. {\bf (g) Rotational feedback cooling:} Levitating the particle in a linearly polarised tweezer and detecting its motion by measuring the scattered light enables feedback cooling via applying a torque through additional cooling lasers \cite{bang2020,vanderlaan2020b}. {\bf (h) Spin cooling:} An asymmetric charged particle with embedded NV centres aligns with the electric quadrupole field generated by the Paul trap. When the spins are pumped into a magnetic state, the B-field exerts a torque, slightly displacing the equilibrium configuration. Driving the spins on the red sideband has been used to cool the nanoparticle librations \cite{delord2020}.}\label{fig:Fig1}
\end{figure*}

\subsection{Reaching the regime of quantum rotations}

\vspace{2mm}

\noindent {\bf Cavity cooling---} Cooling nanorotors to ultra-low rotational temperatures is a prerequisite for many quantum applications. The quantum regime in the centre-of-mass motion of spherical particles has been reached recently \cite{delic2020} by using the techniques of cavity cooling \cite{chang2010,romeroisart2010,barker2010,kiesel2013,asenbaum2013,millen2015,fonseca2016} in the coherent-scattering setup \cite{ritschnjp2009,gonzalezballestero2019,windey2019,delic2019}. Similar techniques can be used for cooling rotations. Illuminating an aspherical particle with a laser induces a position- and orientation-dependent polarization field {inside the particle}, which can be coupled to a high-finesse optical resonator by placing the particle between the cavity mirrors. The laser-induced polarization drives the almost unpopulated cavity mode, whose retarded reaction on the nanoparticle rotation can cool its rotational dynamics into the quantum regime \cite{schafer2020}, see Fig.\,\ref{fig:Fig1}(f).

For linearly polarized laser beams and particles much smaller than the laser wavelength, the rotations around the polarization axis remain undamped. Using elliptically polarized tweezer traps offers a promising strategy for cooling the librations of fully aligned particles. By balancing {shot-noise} heating against the achievable cooling rates, elliptic coherent scattering holds the prospects of simultaneously preparing the rotational and translational motion in the combined 6D quantum ground state \cite{schafer2020}.

\vspace{2mm}

\begin{figure*}
\hrule

\vspace{2mm}

\section*{Box 3: Rotational decoherence}
\justify
{Rotational d}ecoherence {and heating} of nanoparticles due to their interaction with ambient environments is an important limiting factor for quantum superposition experiments \cite{schlosshauer2019}. Common decoherence channels for levitated rotors are collisions with residual gas atoms, scattering of tweezer or black-body photons, thermal emission of radiation, and {phase averaging due to fluctuating stray fields}.

\phantom{xx}
Rotational decoherence of microscale particles may be effectively described by a Markovian quantum master equation of Lindblad form \cite{stickler2016b,zhong2016}. The ensuing jump operators can be related to the microscopic scattering amplitudes for individual collisions between the particle and environmental agents. The rotational motion of nano- to microscale particles is typically much slower than the collisions, so that the environment mainly gains information about the rotor orientation, rather than its angular momentum. The master equation thus describes predominantly how orientational superpositions decay on a timescale {determined by the rotor-environment interaction \cite{papendell2017,pedernales2020b} and} bounded by the scattering rate, see {below}.

\phantom{xx}
In the limit of frequent but weak environmental collisions, the master equation describes angular momentum diffusion  \cite{papendell2017,zhong2017,seberson2020distribution}, as {has been} observed due to  shot noise {in the photon polarization} \cite{vanderlaan2020b}. This can be generalized to the rotational version of quantum Brownian motion, describing rotational friction and thermalization of quantum rotors \cite{stickler2018a}.

\centering
\includegraphics[width = 0.65\textwidth]{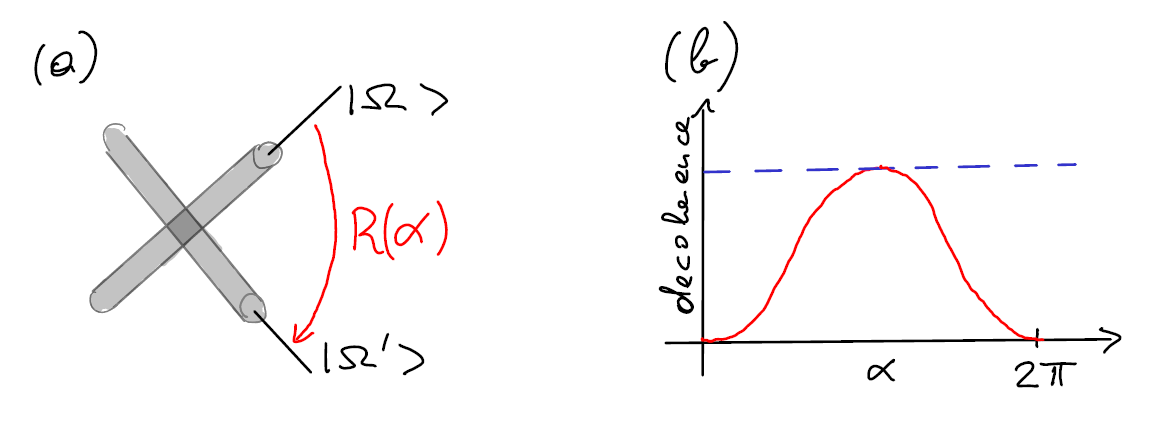}

\justify
{(a) 
Environmental interactions gradually turn a quantum superposition of two different orientations $|\Omega\rangle$ and $|\Omega'\rangle$  into a statistical mixture of definite orientations.
The resulting decoherence rate is a function of the set of angles $\alpha$ required to rotate $|\Omega\rangle$ into $|\Omega'\rangle = {\sf R}(\alpha)|\Omega\rangle$. (b) The rate, sketched here for a symmetric rotor, is determined by the orientation dependence of the scattering amplitudes for collisions with environmental agents; it is bounded by the total scattering rate (blue dashed line)}.

\vspace{2mm}
\hrule

\end{figure*}

\noindent{\bf Feedback cooling---} Feedback cooling relies on monitoring accurately the nanoparticle motion and using this information to apply time-dependent forces and torques serving to reduce the kinetic energy. Nanoscale rotations can be observed with high precision by illuminating the particle with a laser and collecting the scattered light, by balanced \cite{bang2020,vanderlaan2020b} or direct \cite{kuhn2015} detection, or by speckle interferometry \cite{delord2020}. The achievable steady-state temperatures are ultimately limited by detection efficiency and by feedback noise \cite{millen2020}. 

State-of-the-art feedback cooling of the centre-of-mass motion achieves phonon occupations in the quantum regime \cite{magrini2020,tebbenjohanns2021quantum}. Rotational feedback cooling \cite{zhong2017,seberson2019} has been recently demonstrated for dumbbell-shaped particles down to sub-Kelvin temperatures \cite{bang2020,vanderlaan2020b}. In the recent experiment reported in Ref.~\cite{bang2020}, balanced homodyning was used in combination with three linearly polarized cooling lasers to simultaneously cool two rotational and three centre-of-mass degrees of freedom, see Fig.\,\ref{fig:Fig1}(g). The librational trapping potential was controlled by modulating the power of the additional lasers with the particle rotation frequency, achieving librational temperatures below 10\,K. Sub-Kelvin rotational temperatures have been achieved in Ref.\,\cite{vanderlaan2020b}, where angular momentum diffusion \cite{stickler2016b} due to photon shot noise  {in the photon polarization} of a linearly polarized laser was observed. Even lower temperatures are within reach by improving detection.

\vspace{2mm}

\noindent {\bf Spin cooling---} An alternative approach to rotational cooling is to couple the nanoparticle librations to an internal quantum system comprised of a few energy levels and to cool the rotations via driving the internal dynamics. Such schemes exploit that the force or torque acting on the particle depend on the internal state. Similar strategies have been proposed for cooling the centre-of-mass motion \cite{yin2013,pflanzer2013}.

Rotational cooling via few-level systems was recently implemented with charged microscale diamonds, containing approximately $10^9$ nitrogen-vacancy (NV) color centers \cite{delord2017b,delord2017c,delord2018,delord2020}. In these experiments, the particles were levitated in a Paul trap and aligned using their quadrupole moments \cite{delord2017b}, see Fig.\,\ref{fig:Fig1}(h). If the NV spins are prepared in one of their magnetic states, an external magnetic field exerts a torque, slightly shifting the equilibrium orientation of the particle. Driving the nonmagnetic-magnetic NV transition with microwaves on the red sideband allows cooling the rotations of the diamond to a rotational temperature of a few Kelvin  \cite{delord2020}.

\vspace{2mm}

\noindent {\bf Prospects---} Given the recent achievement of preparing a nanoparticle in its centre-of-mass ground state, optical cavity cooling and  feedback cooling seem to be the most promising strategies for reaching the rotational quantum regime in the near future.  {In optical setups}, photon absorption leads to internal heating and therefore to black-body decoherence, see Box 3{, which may prevent subsequent quantum experiments}. {In contrast}, electrical feedback techniques or cooling via embedded spin degrees of freedom avoid {strong laser fields}. {However, anomalous heating due to patch potentials might become relevant for miniaturized electric traps. While magnetic techniques hold the promise of providing ultra-low noise environments \cite{pratcamps2017},} {rotational} heating {due to stray fields coupling to electric and magnetic multipole moments} are not yet fully understood. {In any case, rotational cooling in electric and magnetic traps will require simultaneous centre-of-mass cooling due to coupling between rotations and translations.}

The optimal strategy for rotational cooling may well involve a combination of the cooling and manipulation techniques described above. {While such} hybrid setups have been successfully used for centre-of-mass cooling \cite{millen2015,fonseca2016,tebbenjohanns2019,bykov2019,dania2021}, {rotational cooling schemes need to deal with the intrinsic non-linearity of rotational motion, requiring simultaneous cooling of all three orientational degrees of freedom \cite{seberson2019}.} {For this end, hybrid schemes could} exploit that electric and magnetic moments provide additional handles through which the nanoparticle rotation can be controlled \cite{rider2016}. For instance, combining conservative trapping potentials with  non-conservative torques and with a rotational cooling scheme could enable the preparation of rapidly spinning particles in a state of well-defined angular momentum.

\section{Quantum rotations}

The non-linearity and non-harmonicity of free rigid-body rotations give rise to pronounced quantum effects with no classical analogues -- see Box 2.  In the following, we examine three ideas to observe and exploit such rotational quantum phenomena in an as-of-yet unexplored parameter regime, offering great opportunities for quantum superposition experiments and coherent sensing.

\begin{figure*}
\hrule
\vspace{2mm}

\section*{Box 4: Rotational quantum effects}

\begin{minipage}{0.64\textwidth}
\justify
{\bf Orientational quantum revivals} refer to the torque-free recurrence of an arbitrary rotor state due to the fundamental quantization of angular momentum. Full revivals occur for planar and linear tops, while fractional (partial) revivals also exist for symmetric and asymmetric rotors. The ensuing revival timescale is determined only by the particle moment of inertia and Planck's constant. The observation of orientational quantum revivals with nanoscale particles will enable macroscopic quantum superposition tests and interferometric torque sensing with unprecedented accuracy \cite{stickler2018b}.
\end{minipage}
\begin{minipage}{0.34\textwidth}
\begin{flushright}
\includegraphics[width = 0.85\textwidth]{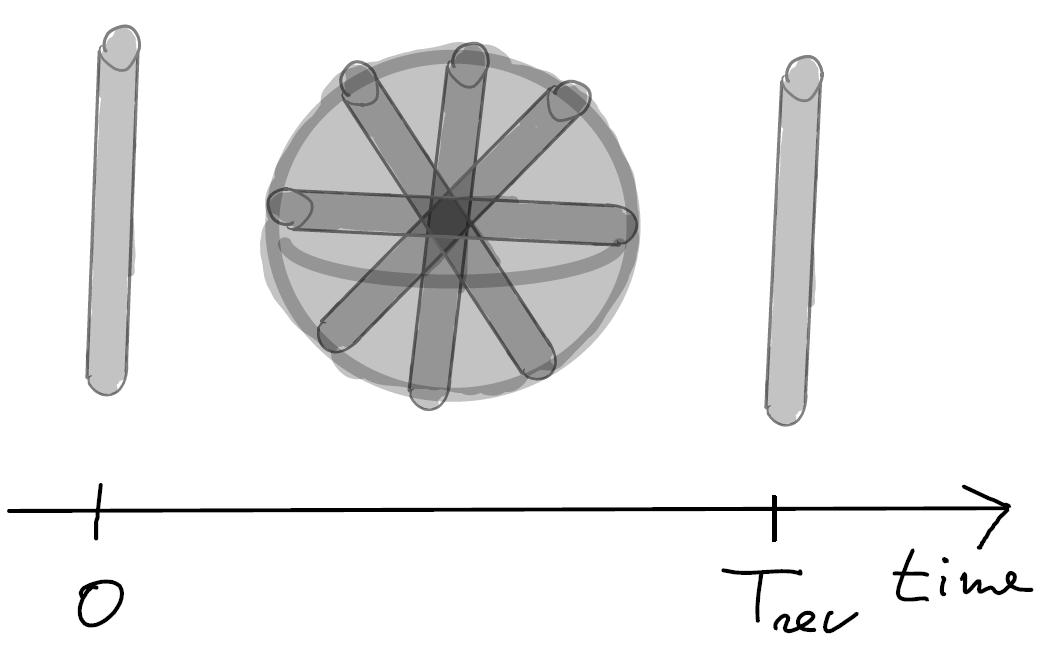}
\end{flushright}
\end{minipage}

\vspace{2mm}

\noindent
\begin{minipage}{0.35\textwidth}
\begin{flushleft}
\includegraphics[width = 0.85\textwidth]{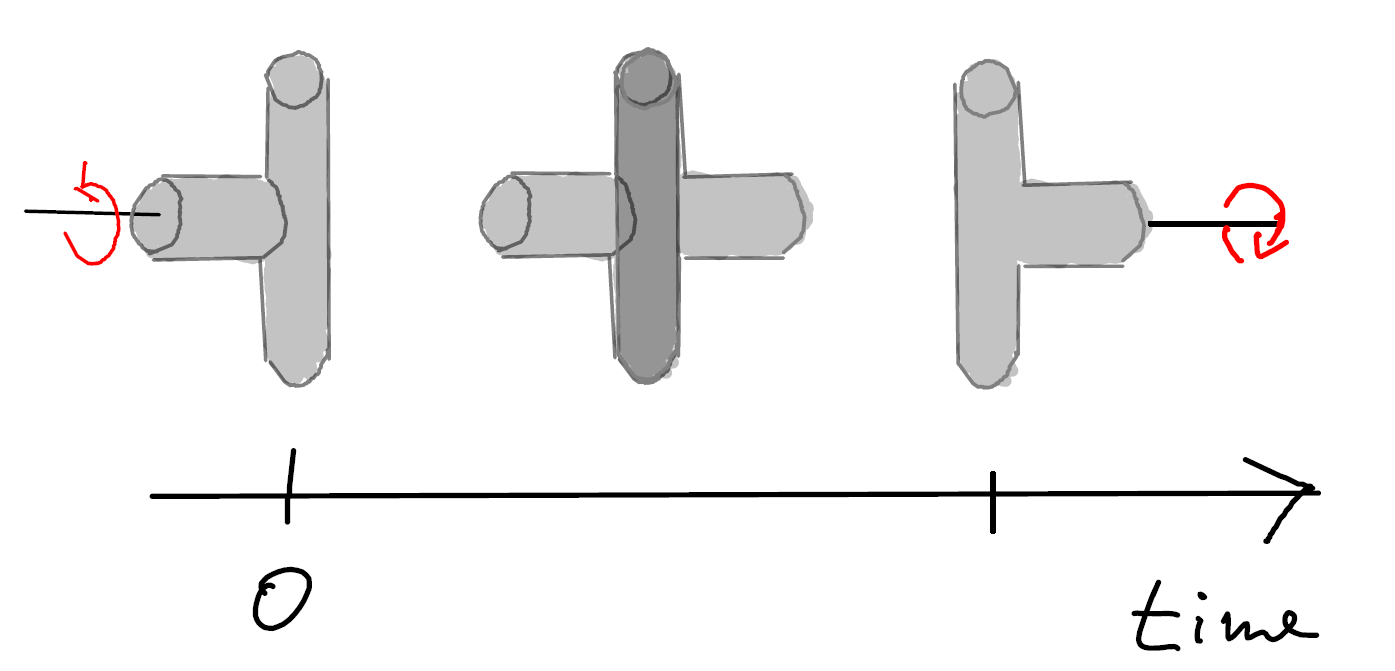}
\end{flushleft}
\end{minipage}
\begin{minipage}{0.64\textwidth}
\justify
Tennis-racket flips (see Box 2) of an ensemble of initial rotation states turn persistent in the quantum regime \cite{ma2020}, a phenomenon dubbed {\bf quantum tennis-racket effect}. Rotors with slightly distinct initial rotation rates exhibit a large range of tennis-racket flipping periods due to the mid-axis instability. Quantum tunnelling and above-barrier reflection of classical rotor trajectories can turn the tennis-racket flips persistent for low temperatures and large rotation speeds.
\end{minipage}

\vspace{2mm}

\noindent
\begin{minipage}{0.6\textwidth}
\justify
{\bf Spin-rotational coupling} is a consequence of the quantum mechanical relation between mechanical angular momentum and magnetization or intrinsic spin. Prominent manifestations are the Einstein-de Haas \cite{einstein1915} and Barnett \cite{barnett1915} effects, describing that flipping a spin changes the rotation rate of magnetized bodies and vice versa. For microscale particles, this spin-rotational interaction can be remarkably strong  \cite{rusconi2017a} {and should become observable} with micromagnets above superconductors \cite{gieseler2020b}.
\end{minipage}
\begin{minipage}{0.35\textwidth}
\begin{flushright}
\includegraphics[width = 0.85\textwidth]{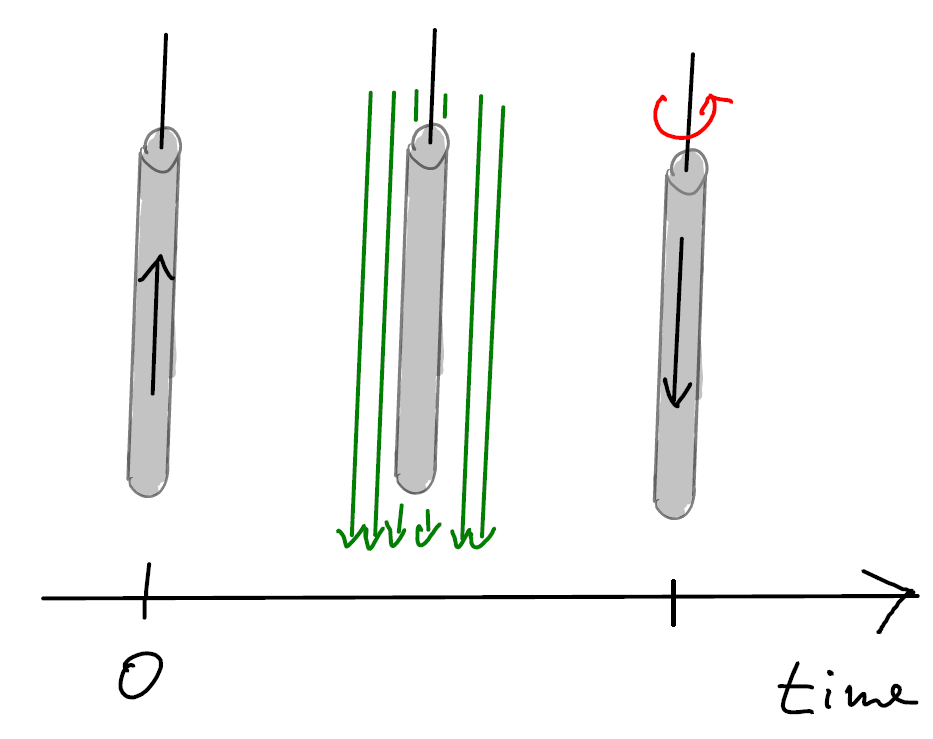}
\end{flushright}
\end{minipage}

\vspace{2mm}

\hrule
\end{figure*}

\subsection{Orientational quantum revivals}\label{sec:revival}
We first discuss  orientational quantum revivals of symmetric nanorotors, see Box 2, a phenomenon expected to be observable using state-of-the art technology \cite{stickler2018b}. Such an experiment will first cool the librations of the trapped rotor to align its symmetry axis with a space-fixed direction, thereby creating a broad superposition of angular momentum states. Upon release, the torque-free dynamics disperses the initially aligned state to almost uniformly distributed rotor orientations. The timescale of the dispersion is dominated by the classical angular momentum distribution or by the quantum uncertainty, depending on the trap stiffness and the temperature of the initial state.

Most of the time, the expected rotor orientation remains close to the classical prediction for the corresponding initial thermal state. However, the dynamics can show pronounced quantum interference phenomena at certain instances, as a consequence of angular momentum quantization. Specifically, the spacing of rotational energies increases linearly with the total angular momentum quantum number. This implies that the relative phases between all angular momentum states vanish at integer multiples of the {\it quantum revival time} $T_{\rm rev} = 2\pi I/\hbar$,  determined by the moment of inertia $I$ for rotations orthogonal to the symmetry axis. An initially aligned rotor thus perfectly returns to its initial orientation -- a pronounced quantum interference effect with no classical analogue. 

The revival time grows with the fifth power of the rotor length, quickly reaching cosmological time spans for macroscopic objects. For instance, even a $10$\,micron long silicon cylinder with diameter of $100$\,nanometers shows a revival time of  $10^8$ seconds, i.e. more than $3$ years. In contrast,  quantum revivals are expected already after a few milli-seconds for a $50$\,nanometer-sized silicon nanorotor or carbon nanotube. Such coherence times are realistically achievable in the presence of environmentally induced rotational decoherence, see Box 3.

The experiment proposed in Ref.~\cite{stickler2018b} consists of four steps: First, the rotor is trapped in a linearly polarized optical tweezer and its librations are cooled to sub-Kelvin temperatures {so that its symmetry axis is strongly aligned with the field polarization}. Switching off the laser releases the particle. During the free fall of the centre of mass, at a pressure of less than $10^{-8}$\,mbar, the orientation state quickly disperses, remaining uniformly distributed for most of the time.  After the quantum revival time, amounting to a few milli-seconds, the rotor briefly flashes back to its initial alignment {for a few microseconds}. The duration of this recurrence is given by the dispersion timescale and thus determined by the temperature of the aligned initial state{, with lower temperatures giving rise to longer revivals}. The revival can be observed by optical light scattering, which measures the relative orientation between the rotor axis and the polarization of the probe beam. After measurement, the tweezer is switched on again to recapture the rotor, which is then recycled for the next experimental run.

\subsection{Quantum tennis-racket flips}\label{sec:flips}
We next consider  quantum-persistent tennis-racket rotations, see Box 2, of asymmetric rotors. The latter are characterized by three distinct moments of inertia associated with three body-fixed principal axes. While classical rotations around the axes of maximum and minimum inertia are stable, revolutions around the mid-axis are always unstable \cite{goldstein}, see Box 2. The ensuing {\it tennis-racket} effect describes the periodic flipping of the mid-axis orientation at a rate set by the initial conditions. For initial conditions approaching perfect mid-axis rotations, the period of tennis-racket flips grows exponentially. Consequently, tennis-racket flips quickly average out if the initial angular momentum is not precisely controlled.

In the quantum realm, quantum tunnelling and above-barrier reflection blur the distinction between semiclassical trajectories associated with tennis-racket flips via major-axis rotations and minor-axis rotations \cite{ma2020}. In particular, perfect mid-axis rotations with angular momentum only in direction of the mid-axis are prevented by the quantum uncertainty of the angular-momentum vector. This leads to an increased energy spacing in the vicinity of mid-axis rotations, rendering tennis racket flips persistent. The impact of tunnelling and above-barrier reflection increases with the angular velocity $\omega$ of the particle. This implies that quantum tennis-racket flips can be observed provided $\hbar \omega/ k_{\rm B} T \gtrsim 0.1$, where the temperature $T$ quantifies the angular momentum width.

A possible experiment to observe this effect consists of the following steps: the rotor mid axis is first aligned and cooled to {milli-Kelvin} temperatures. The mid-axis is then accelerated, for instance by light scattering \cite{schafer2020}, until {GHz rotation rates} are reached. Upon reaching {this} rotation rate, all fields are switched off and the rotor is left to tumble freely. As in the quantum-revival experiment, the rotor orientation is observed via light scattering from a weak probe pulse {after several microseconds}, and  re-trapping enables repeated measurements  \cite{ma2020}. {If the rotor were initially cooled to $\mu$K temperatures, MHz rotation rates would suffice to observe the effect on millisecond timescales. Pressures lower than $10^{-8}$\,mbar are required to suppress  decoherence due to gas scattering.}

\subsection{Spin-stabilised levitation}
The magnetization of an object can be converted into its mechanical rotation and vice versa -- see Box 2. These interconversions, known as the Einstein-de Haas \cite{einstein1915} and the Barnett \cite{barnett1915} effects, can be significant for microscale particles, even though they have only  little influence on the rotations of {more massive} objects. For instance, flipping a single spin makes a nanometer-sized Cobalt sphere rotate with  MHz frequency \cite{rusconi2017a}.

Macroscopic electrodynamics describes magnetization as caused by microscopic loop currents of charged matter \cite{jackson1999}. 
The associated angular momentum contributes to the total angular momentum of any magnetized object according to the gyromagnetic ratio of the material, even if the magnetization is due to spin. The conservation of total angular momentum then implies that mechanical rotation can be induced without applying a torque, only by magnetizing the body. This is known as the Einstein-de Haas effect. Vice versa, spinning a neutral object induces a magnetization field, known as the Barnett effect. As as a consequence, the free rotation dynamics of magnetic particles can strongly deviate from the dynamics of dielectric rotors \cite{rusconi2016,kimball2016,rusconi2017a}.

Spin-rotational coupling enables the stable levitation of a non-rotating magnet in a magneto-static field \cite{rusconi2017a,rusconi2017b}. The effect thus  circumvents Earnshaw's theorem, which states that it is impossible to levitate a stationary charge or current distribution in free space by means of static electro-magnetic
fields \cite{jackson1999}. It was therefore believed that stable trapping of electric charges or magnets requires oscillating fields, as exploited in quadrupole ion traps, or the rotation of the levitated object, as used in the {\it Levitron} \cite{berry1996,simon1997spin}. Yet,  levitation can be achieved even with static fields and stationary objects since the magnetization is equivalent to mechanical rotation \cite{rusconi2016,rusconi2017a,rusconi2017b}. A detailed analysis of the combined rotation-translational dynamics of a {nanometer-sized Cobalt sphere} in a Ioffe-Pritchard  {magneto-static} field {of milli-Tesla field strength} shows that already a single spin suffices for {stable levitation} \cite{rusconi2017a}.

{
\subsection{Challenges}
The observation of quantum rotations with nanoparticles faces a number of technological obstacles in addition to implementing rotational cooling and mitigating decoherence. First, rotational interference experiments depend sensitively on the inertia tensor, implying that a precise knowledge of the particle size and shape is required. For instance, the revival time of a nanoscale rod increases with the third power of its length. In principle, the particle geometry can be extracted in-situ from the power spectra of the light scattered off a deeply trapped nanorotor \cite{kuhn2017a,vanderlaan2020}. Even then, the strong shape dependence of rotational interference necessitates  quantum experiments to be performed with a single well-characterized particle, which is recycled after each experimental run until the interference pattern becomes visible. Second, observing rotational interference, as discussed in the proposals Secs.~\ref{sec:revival} and \ref{sec:flips}, requires detecting the rotor orientation after a period of free revolution. This should be possible by light scattering from a weak probe pulse, but requires detecting intensity variations on top of a potentially large background. An alternative strategy may be projective measurements of embedded paramagnetic impurities, such as NV centres, whose quantization axis is tied to the body-fixed frame.}

{On the theory side, it is challenging to describe quantum rotations in the limit that billions of total angular momentum states are involved. This is quickly the case  for asymmetric nanoscale particles, whose initial state is strongly aligned in a trap.  Given that exact numerical diagonalisation is impossible due to the vast Hilbert space dimension (growing as the cube of the total angular momentum), effective semiclassical descriptions are required. Such theories must account for the discreteness in the relevant parts of the spectrum while approximating others as continuous. This will also involve formulating suitable stochastic unravellings to simulate the decohering dynamics of nanorotors in presence of realistic environments. In this regard, the impact of stray electric fields is not yet sufficiently well understood, as are magnetization-induced sources of rotational heating. Finally, experiments with microscale rotors will require a systematic description of optical torques and light scattering beyond the dipole approximation \cite{stickler2016a,seberson2020}.}

\section{Future Applications}

Once nanorotors have been prepared in the quantum regime {and all technological challenges have been overcome}, their dynamics can be exploited for technology and tests of fundamental physics. They will improve  state-of-the-art torque and rotation sensing by orders of magnitude, and contribute significantly to the exploration of force laws at short distances, the search for physics beyond the Standard Model, and tests of wavefunction collapse models.

{{\bf Sensing---}} Torque sensing at and beyond the standard quantum limit  can be achieved by applying established quantum optomechanical sensing schemes \cite{bowen2015} to the librations of a deeply trapped nanorotor \cite{hoang2016,ahn2020,schafer2020}. Levitating the particle close to a surface will enable measurements of the Casimir torque \cite{xu2017}. {Provided the Casimir torque has been observed, r}otational vacuum friction {might become} observable \cite{ahn2020}, an as-of-yet unobserved damping {torque} due to material-induced retardation of the Casimir-Polder interaction \cite{zhao2012}. {These experiments might also enable measuring short-range interactions due to the charge distribution of the nanoparticle surface, thus providing an ideal testbed for studying charge-induced rotational decoherence and heating due to fluctuating polarization densities at the surface.} Centre-of-mass explorations of force laws at short distances \cite{moore2020} can be readily extended to the librations of aspherical particles. This might enable simultaneous rotation-translational measurement schemes that are sensitive to the direction of the force fields. {The precession of levitated ferromagnetic rotors can in principle be used for quantum-limited measurements of external magnetic fields via observation of the macrospin \cite{kimball2016,band2018}. However, the isolation required for such magnetometers renders their implementation challenging.}

{Static t}orques can also be measured via their effect on rotational superposition states. In analogy to atom interferometric sensors, such schemes would exploit that the rotational interference signal, such as the revival intensity, depend sensitively on the presence of external torques. Sensitivity to static electro-magnetic fields can be achieved by using polar or magnetized particles. For instance, observing the quantum revival of a $50$\,nm linear rotor with a single charge at its endpoint would enable sensing fields on the order of ${\rm mV/m}$. This amounts to torques on the order of $10^{-30}$\,N\,m,  allowing 3D mapping of the local field lines\cite{stickler2018b}. {The feasibility of such interferometric sensing schemes will depend strongly on charge-induced decoherence due to e.g. stray electric fields (see Box 3).}

{{\bf Fundamental tests---}} The verification of rotational quantum superpositions would also serve to test models of an objective wavefunction collapse. Such theories modify quantum mechanics with the aim of recovering macro-realism without contradicting experimental observations. A prominent example is {\it Continuous Spontaneous Localization} \cite{bassi2013}, predicting a universal spatial and orientational decoherence and heating mechanism \cite{schrinski2017}. Observing quantum interference or the lack of collapse-induced heating rules out combinations of the parameters characterising the model \cite{schrinski2017,carlesso2018a}. The shape-dependent centre-of-mass and rotational heating rates can be measured with a single trapped particle, facilitating absolute measurements of the collapse parameters.

{Finally, p}recisely monitoring the dynamics of magnetic rotors might probe for anomalous spin-spin interactions proposed by certain axion models \cite{fadeev2021ferromagnetic}, the Lense-Thirring effect on magnetized objects \cite{fadeev2021gravity}, or simulate many-body dynamics with embedded NV centers \cite{ma2016}. {Rotation states} might also be useful for quantum information processing \cite{grimsmo2020,albert2020}{, even though the feasibility of these proposals with nanoparticles is unlikely.}

{{\bf Conclusion---} Q}uantum rotations of microscale particles offer a promising route into an unexplored regime of quantum physics. The recent developments bring the quantum regime and the first observation of microscale rotational quantum interference within reach. These effects are manifestations of the non-linearity of  rotations, with no  analogue in  centre-of-mass dynamics. Their observation would constitute a test of the quantum superposition principle, achieving a degree of macroscopicity \cite{schrinski2019} on a par with {center-of-mass} matter-wave interference. Rotational superpositions may thus offer a viable path towards observing and exploiting the most macroscopic mechanical Schrödinger cat states.

\begin{acknowledgements}
The authors thank James Millen and Cosimo Rusconi for comments on the manuscript. BAS acknowledges funding from the Deutsche Forschungsgemeinschaft (DFG, German Research Foundation) -- 411042854. KH acknowledges funding from the Deutsche Forschungsgemeinschaft (DFG, German Research Foundation) -- 394398290. MSK was supported by the QuantERA ERA-NET Cofund in Quantum Technologies implemented within the European Union's Horizon 2020 Programme.
\end{acknowledgements}


\begin{thebibliography}{104}%
\makeatletter
\providecommand \@ifxundefined [1]{%
 \@ifx{#1\undefined}
}%
\providecommand \@ifnum [1]{%
 \ifnum #1\expandafter \@firstoftwo
 \else \expandafter \@secondoftwo
 \fi
}%
\providecommand \@ifx [1]{%
 \ifx #1\expandafter \@firstoftwo
 \else \expandafter \@secondoftwo
 \fi
}%
\providecommand \natexlab [1]{#1}%
\providecommand \enquote  [1]{``#1''}%
\providecommand \bibnamefont  [1]{#1}%
\providecommand \bibfnamefont [1]{#1}%
\providecommand \citenamefont [1]{#1}%
\providecommand \href@noop [0]{\@secondoftwo}%
\providecommand \href [0]{\begingroup \@sanitize@url \@href}%
\providecommand \@href[1]{\@@startlink{#1}\@@href}%
\providecommand \@@href[1]{\endgroup#1\@@endlink}%
\providecommand \@sanitize@url [0]{\catcode `\\12\catcode `\$12\catcode
  `\&12\catcode `\#12\catcode `\^12\catcode `\_12\catcode `\%12\relax}%
\providecommand \@@startlink[1]{}%
\providecommand \@@endlink[0]{}%
\providecommand \url  [0]{\begingroup\@sanitize@url \@url }%
\providecommand \@url [1]{\endgroup\@href {#1}{\urlprefix }}%
\providecommand \urlprefix  [0]{URL }%
\providecommand \Eprint [0]{\href }%
\providecommand \doibase [0]{http://dx.doi.org/}%
\providecommand \selectlanguage [0]{\@gobble}%
\providecommand \bibinfo  [0]{\@secondoftwo}%
\providecommand \bibfield  [0]{\@secondoftwo}%
\providecommand \translation [1]{[#1]}%
\providecommand \BibitemOpen [0]{}%
\providecommand \bibitemStop [0]{}%
\providecommand \bibitemNoStop [0]{.\EOS\space}%
\providecommand \EOS [0]{\spacefactor3000\relax}%
\providecommand \BibitemShut  [1]{\csname bibitem#1\endcsname}%
\let\auto@bib@innerbib\@empty
\bibitem [{\citenamefont {Goldstein}(1980)}]{goldstein}%
  \BibitemOpen
  \bibfield  {author} {\bibinfo {author} {\bibfnamefont {H.}~\bibnamefont
  {Goldstein}},\ }\href@noop {} {\emph {\bibinfo {title} {{Classical
  Mechanics}}}}\ (\bibinfo  {publisher} {Addison-Wesley Publishing Company -
  Reading, Massachusetts},\ \bibinfo {year} {1980})\BibitemShut {NoStop}%
\bibitem [{\citenamefont {Millen}\ \emph {et~al.}(2020)\citenamefont {Millen},
  \citenamefont {Monteiro}, \citenamefont {Pettit},\ and\ \citenamefont
  {Vamivakas}}]{millen2020}%
  \BibitemOpen
  \bibfield  {author} {\bibinfo {author} {\bibfnamefont {J.}~\bibnamefont
  {Millen}}, \bibinfo {author} {\bibfnamefont {T.~S.}\ \bibnamefont
  {Monteiro}}, \bibinfo {author} {\bibfnamefont {R.}~\bibnamefont {Pettit}}, \
  and\ \bibinfo {author} {\bibfnamefont {A.~N.}\ \bibnamefont {Vamivakas}},\
  }\bibfield  {title} {\emph {\bibinfo {title} {Optomechanics with levitated
  particles},\ }}\href@noop {} {\bibfield  {journal} {\bibinfo  {journal} {Rep.
  Prog. Phys.}\ }\textbf {\bibinfo {volume} {83}},\ \bibinfo {pages} {026401}
  (\bibinfo {year} {2020})}\BibitemShut {NoStop}%
\bibitem [{\citenamefont {Hoang}\ \emph {et~al.}(2016)\citenamefont {Hoang},
  \citenamefont {Ma}, \citenamefont {Ahn}, \citenamefont {Bang}, \citenamefont
  {Robicheaux}, \citenamefont {Yin},\ and\ \citenamefont {Li}}]{hoang2016}%
  \BibitemOpen
  \bibfield  {author} {\bibinfo {author} {\bibfnamefont {T.~M.}\ \bibnamefont
  {Hoang}}, \bibinfo {author} {\bibfnamefont {Y.}~\bibnamefont {Ma}}, \bibinfo
  {author} {\bibfnamefont {J.}~\bibnamefont {Ahn}}, \bibinfo {author}
  {\bibfnamefont {J.}~\bibnamefont {Bang}}, \bibinfo {author} {\bibfnamefont
  {F.}~\bibnamefont {Robicheaux}}, \bibinfo {author} {\bibfnamefont {Z.-Q.}\
  \bibnamefont {Yin}}, \ and\ \bibinfo {author} {\bibfnamefont
  {T.}~\bibnamefont {Li}},\ }\bibfield  {title} {\emph {\bibinfo {title}
  {Torsional optomechanics of a levitated nonspherical nanoparticle},\
  }}\href@noop {} {\bibfield  {journal} {\bibinfo  {journal} {Phys. Rev.
  Lett.}\ }\textbf {\bibinfo {volume} {117}},\ \bibinfo {pages} {123604}
  (\bibinfo {year} {2016})}\BibitemShut {NoStop}%
\bibitem [{\citenamefont {Kuhn}\ \emph
  {et~al.}(2017{\natexlab{a}})\citenamefont {Kuhn}, \citenamefont {Kosloff},
  \citenamefont {Stickler}, \citenamefont {Patolsky}, \citenamefont
  {Hornberger}, \citenamefont {Arndt},\ and\ \citenamefont
  {Millen}}]{kuhn2017a}%
  \BibitemOpen
  \bibfield  {author} {\bibinfo {author} {\bibfnamefont {S.}~\bibnamefont
  {Kuhn}}, \bibinfo {author} {\bibfnamefont {A.}~\bibnamefont {Kosloff}},
  \bibinfo {author} {\bibfnamefont {B.~A.}\ \bibnamefont {Stickler}}, \bibinfo
  {author} {\bibfnamefont {F.}~\bibnamefont {Patolsky}}, \bibinfo {author}
  {\bibfnamefont {K.}~\bibnamefont {Hornberger}}, \bibinfo {author}
  {\bibfnamefont {M.}~\bibnamefont {Arndt}}, \ and\ \bibinfo {author}
  {\bibfnamefont {J.}~\bibnamefont {Millen}},\ }\bibfield  {title} {\emph
  {\bibinfo {title} {Full rotational control of levitated silicon nanorods},\
  }}\href@noop {} {\bibfield  {journal} {\bibinfo  {journal} {Optica}\ }\textbf
  {\bibinfo {volume} {4}},\ \bibinfo {pages} {356} (\bibinfo {year}
  {2017}{\natexlab{a}})}\BibitemShut {NoStop}%
\bibitem [{\citenamefont {Kuhn}\ \emph
  {et~al.}(2017{\natexlab{b}})\citenamefont {Kuhn}, \citenamefont {Stickler},
  \citenamefont {Kosloff}, \citenamefont {Patolsky}, \citenamefont
  {Hornberger}, \citenamefont {Arndt},\ and\ \citenamefont
  {Millen}}]{kuhn2017b}%
  \BibitemOpen
  \bibfield  {author} {\bibinfo {author} {\bibfnamefont {S.}~\bibnamefont
  {Kuhn}}, \bibinfo {author} {\bibfnamefont {B.~A.}\ \bibnamefont {Stickler}},
  \bibinfo {author} {\bibfnamefont {A.}~\bibnamefont {Kosloff}}, \bibinfo
  {author} {\bibfnamefont {F.}~\bibnamefont {Patolsky}}, \bibinfo {author}
  {\bibfnamefont {K.}~\bibnamefont {Hornberger}}, \bibinfo {author}
  {\bibfnamefont {M.}~\bibnamefont {Arndt}}, \ and\ \bibinfo {author}
  {\bibfnamefont {J.}~\bibnamefont {Millen}},\ }\bibfield  {title} {\emph
  {\bibinfo {title} {Optically driven ultra-stable nanomechanical rotor},\
  }}\href@noop {} {\bibfield  {journal} {\bibinfo  {journal} {Nat. Commun.}\
  }\textbf {\bibinfo {volume} {8}},\ \bibinfo {pages} {1670} (\bibinfo {year}
  {2017}{\natexlab{b}})}\BibitemShut {NoStop}%
\bibitem [{\citenamefont {Rashid}\ \emph {et~al.}(2018)\citenamefont {Rashid},
  \citenamefont {Toro{\v{s}}}, \citenamefont {Setter},\ and\ \citenamefont
  {Ulbricht}}]{rashid2018}%
  \BibitemOpen
  \bibfield  {author} {\bibinfo {author} {\bibfnamefont {M.}~\bibnamefont
  {Rashid}}, \bibinfo {author} {\bibfnamefont {M.}~\bibnamefont {Toro{\v{s}}}},
  \bibinfo {author} {\bibfnamefont {A.}~\bibnamefont {Setter}}, \ and\ \bibinfo
  {author} {\bibfnamefont {H.}~\bibnamefont {Ulbricht}},\ }\bibfield  {title}
  {\emph {\bibinfo {title} {Precession motion in levitated optomechanics},\
  }}\href@noop {} {\bibfield  {journal} {\bibinfo  {journal} {Phys. Rev.
  Lett.}\ }\textbf {\bibinfo {volume} {121}},\ \bibinfo {pages} {253601}
  (\bibinfo {year} {2018})}\BibitemShut {NoStop}%
\bibitem [{\citenamefont {Reimann}\ \emph {et~al.}(2018)\citenamefont
  {Reimann}, \citenamefont {Doderer}, \citenamefont {Hebestreit}, \citenamefont
  {Diehl}, \citenamefont {Frimmer}, \citenamefont {Windey}, \citenamefont
  {Tebbenjohanns},\ and\ \citenamefont {Novotny}}]{reimann2018}%
  \BibitemOpen
  \bibfield  {author} {\bibinfo {author} {\bibfnamefont {R.}~\bibnamefont
  {Reimann}}, \bibinfo {author} {\bibfnamefont {M.}~\bibnamefont {Doderer}},
  \bibinfo {author} {\bibfnamefont {E.}~\bibnamefont {Hebestreit}}, \bibinfo
  {author} {\bibfnamefont {R.}~\bibnamefont {Diehl}}, \bibinfo {author}
  {\bibfnamefont {M.}~\bibnamefont {Frimmer}}, \bibinfo {author} {\bibfnamefont
  {D.}~\bibnamefont {Windey}}, \bibinfo {author} {\bibfnamefont
  {F.}~\bibnamefont {Tebbenjohanns}}, \ and\ \bibinfo {author} {\bibfnamefont
  {L.}~\bibnamefont {Novotny}},\ }\bibfield  {title} {\emph {\bibinfo {title}
  {{GHz} rotation of an optically trapped nanoparticle in vacuum},\ }}\href
  {\doibase 10.1103/PhysRevLett.121.033602} {\bibfield  {journal} {\bibinfo
  {journal} {Phys. Rev. Lett.}\ }\textbf {\bibinfo {volume} {121}},\ \bibinfo
  {pages} {033602} (\bibinfo {year} {2018})}\BibitemShut {NoStop}%
\bibitem [{\citenamefont {Ahn}\ \emph {et~al.}(2018)\citenamefont {Ahn},
  \citenamefont {Xu}, \citenamefont {Bang}, \citenamefont {Deng}, \citenamefont
  {Hoang}, \citenamefont {Han}, \citenamefont {Ma},\ and\ \citenamefont
  {Li}}]{ahn2018}%
  \BibitemOpen
  \bibfield  {author} {\bibinfo {author} {\bibfnamefont {J.}~\bibnamefont
  {Ahn}}, \bibinfo {author} {\bibfnamefont {Z.}~\bibnamefont {Xu}}, \bibinfo
  {author} {\bibfnamefont {J.}~\bibnamefont {Bang}}, \bibinfo {author}
  {\bibfnamefont {Y.-H.}\ \bibnamefont {Deng}}, \bibinfo {author}
  {\bibfnamefont {T.~M.}\ \bibnamefont {Hoang}}, \bibinfo {author}
  {\bibfnamefont {Q.}~\bibnamefont {Han}}, \bibinfo {author} {\bibfnamefont
  {R.-M.}\ \bibnamefont {Ma}}, \ and\ \bibinfo {author} {\bibfnamefont
  {T.}~\bibnamefont {Li}},\ }\bibfield  {title} {\emph {\bibinfo {title}
  {Optically levitated nanodumbbell torsion balance and {GHz} nanomechanical
  rotor},\ }}\href {https://link.aps.org/doi/10.1103/PhysRevLett.121.033603}
  {\bibfield  {journal} {\bibinfo  {journal} {Phys. Rev. Lett.}\ }\textbf
  {\bibinfo {volume} {121}},\ \bibinfo {pages} {033603} (\bibinfo {year}
  {2018})}\BibitemShut {NoStop}%
\bibitem [{\citenamefont {Jin}\ \emph {et~al.}(2020)\citenamefont {Jin},
  \citenamefont {Yan}, \citenamefont {Rahman}, \citenamefont {Li},
  \citenamefont {Yu},\ and\ \citenamefont {Zhang}}]{jin2020}%
  \BibitemOpen
  \bibfield  {author} {\bibinfo {author} {\bibfnamefont {Y.}~\bibnamefont
  {Jin}}, \bibinfo {author} {\bibfnamefont {J.}~\bibnamefont {Yan}}, \bibinfo
  {author} {\bibfnamefont {S.~J.}\ \bibnamefont {Rahman}}, \bibinfo {author}
  {\bibfnamefont {J.}~\bibnamefont {Li}}, \bibinfo {author} {\bibfnamefont
  {X.}~\bibnamefont {Yu}}, \ and\ \bibinfo {author} {\bibfnamefont
  {J.}~\bibnamefont {Zhang}},\ }\bibfield  {title} {\emph {\bibinfo {title} {6
  {GHz} hyperfast rotation of an optically levitated nanosphere in vacuum},\
  }}\href@noop {} {\bibfield  {journal} {\bibinfo  {journal} {arXiv preprint
  arXiv:2012.09693}\ } (\bibinfo {year} {2020})}\BibitemShut {NoStop}%
\bibitem [{\citenamefont {Ahn}\ \emph {et~al.}(2020)\citenamefont {Ahn},
  \citenamefont {Xu}, \citenamefont {Bang}, \citenamefont {Ju}, \citenamefont
  {Gao},\ and\ \citenamefont {Li}}]{ahn2020}%
  \BibitemOpen
  \bibfield  {author} {\bibinfo {author} {\bibfnamefont {J.}~\bibnamefont
  {Ahn}}, \bibinfo {author} {\bibfnamefont {Z.}~\bibnamefont {Xu}}, \bibinfo
  {author} {\bibfnamefont {J.}~\bibnamefont {Bang}}, \bibinfo {author}
  {\bibfnamefont {P.}~\bibnamefont {Ju}}, \bibinfo {author} {\bibfnamefont
  {X.}~\bibnamefont {Gao}}, \ and\ \bibinfo {author} {\bibfnamefont
  {T.}~\bibnamefont {Li}},\ }\bibfield  {title} {\emph {\bibinfo {title}
  {Ultrasensitive torque detection with an optically levitated nanorotor},\
  }}\href@noop {} {\bibfield  {journal} {\bibinfo  {journal} {Nat. Nanotechn.}\
  }\textbf {\bibinfo {volume} {15}},\ \bibinfo {pages} {89–93} (\bibinfo
  {year} {2020})}\BibitemShut {NoStop}%
\bibitem [{\citenamefont {Deli{\'c}}\ \emph {et~al.}(2020)\citenamefont
  {Deli{\'c}}, \citenamefont {Reisenbauer}, \citenamefont {Dare}, \citenamefont
  {Grass}, \citenamefont {Vuleti{\'c}}, \citenamefont {Kiesel},\ and\
  \citenamefont {Aspelmeyer}}]{delic2020}%
  \BibitemOpen
  \bibfield  {author} {\bibinfo {author} {\bibfnamefont {U.}~\bibnamefont
  {Deli{\'c}}}, \bibinfo {author} {\bibfnamefont {M.}~\bibnamefont
  {Reisenbauer}}, \bibinfo {author} {\bibfnamefont {K.}~\bibnamefont {Dare}},
  \bibinfo {author} {\bibfnamefont {D.}~\bibnamefont {Grass}}, \bibinfo
  {author} {\bibfnamefont {V.}~\bibnamefont {Vuleti{\'c}}}, \bibinfo {author}
  {\bibfnamefont {N.}~\bibnamefont {Kiesel}}, \ and\ \bibinfo {author}
  {\bibfnamefont {M.}~\bibnamefont {Aspelmeyer}},\ }\bibfield  {title} {\emph
  {\bibinfo {title} {Cooling of a levitated nanoparticle to the motional
  quantum ground state},\ }}\href@noop {} {\bibfield  {journal} {\bibinfo
  {journal} {Science}\ }\textbf {\bibinfo {volume} {367}},\ \bibinfo {pages}
  {892} (\bibinfo {year} {2020})}\BibitemShut {NoStop}%
\bibitem [{\citenamefont {Magrini}\ \emph {et~al.}(2020)\citenamefont
  {Magrini}, \citenamefont {Rosenzweig}, \citenamefont {Bach}, \citenamefont
  {Deutschmann-Olek}, \citenamefont {Hofer}, \citenamefont {Hong},
  \citenamefont {Kiesel}, \citenamefont {Kugi},\ and\ \citenamefont
  {Aspelmeyer}}]{magrini2020}%
  \BibitemOpen
  \bibfield  {author} {\bibinfo {author} {\bibfnamefont {L.}~\bibnamefont
  {Magrini}}, \bibinfo {author} {\bibfnamefont {P.}~\bibnamefont {Rosenzweig}},
  \bibinfo {author} {\bibfnamefont {C.}~\bibnamefont {Bach}}, \bibinfo {author}
  {\bibfnamefont {A.}~\bibnamefont {Deutschmann-Olek}}, \bibinfo {author}
  {\bibfnamefont {S.~G.}\ \bibnamefont {Hofer}}, \bibinfo {author}
  {\bibfnamefont {S.}~\bibnamefont {Hong}}, \bibinfo {author} {\bibfnamefont
  {N.}~\bibnamefont {Kiesel}}, \bibinfo {author} {\bibfnamefont
  {A.}~\bibnamefont {Kugi}}, \ and\ \bibinfo {author} {\bibfnamefont
  {M.}~\bibnamefont {Aspelmeyer}},\ }\bibfield  {title} {\emph {\bibinfo
  {title} {Optimal quantum control of mechanical motion at room temperature:
  ground-state cooling},\ }}\href@noop {} {\bibfield  {journal} {\bibinfo
  {journal} {aXiv preprint arXiv:2012.15188}\ } (\bibinfo {year}
  {2020})}\BibitemShut {NoStop}%
\bibitem [{\citenamefont {Tebbenjohanns}\ \emph {et~al.}(2021)\citenamefont
  {Tebbenjohanns}, \citenamefont {Mattana}, \citenamefont {Rossi},
  \citenamefont {Frimmer},\ and\ \citenamefont
  {Novotny}}]{tebbenjohanns2021quantum}%
  \BibitemOpen
  \bibfield  {author} {\bibinfo {author} {\bibfnamefont {F.}~\bibnamefont
  {Tebbenjohanns}}, \bibinfo {author} {\bibfnamefont {M.~L.}\ \bibnamefont
  {Mattana}}, \bibinfo {author} {\bibfnamefont {M.}~\bibnamefont {Rossi}},
  \bibinfo {author} {\bibfnamefont {M.}~\bibnamefont {Frimmer}}, \ and\
  \bibinfo {author} {\bibfnamefont {L.}~\bibnamefont {Novotny}},\ }\bibfield
  {title} {\emph {\bibinfo {title} {Quantum control of a nanoparticle optically
  levitated in cryogenic free space},\ }}\href@noop {} {\bibfield  {journal}
  {\bibinfo  {journal} {arXiv:2103.03853}\ } (\bibinfo {year}
  {2021})}\BibitemShut {NoStop}%
\bibitem [{\citenamefont {Delord}\ \emph {et~al.}(2020)\citenamefont {Delord},
  \citenamefont {Huillery}, \citenamefont {Nicolas},\ and\ \citenamefont
  {H{\'e}tet}}]{delord2020}%
  \BibitemOpen
  \bibfield  {author} {\bibinfo {author} {\bibfnamefont {T.}~\bibnamefont
  {Delord}}, \bibinfo {author} {\bibfnamefont {P.}~\bibnamefont {Huillery}},
  \bibinfo {author} {\bibfnamefont {L.}~\bibnamefont {Nicolas}}, \ and\
  \bibinfo {author} {\bibfnamefont {G.}~\bibnamefont {H{\'e}tet}},\ }\bibfield
  {title} {\emph {\bibinfo {title} {Spin-cooling of the motion of a trapped
  diamond},\ }}\href@noop {} {\bibfield  {journal} {\bibinfo  {journal}
  {Nature}\ }\textbf {\bibinfo {volume} {580}},\ \bibinfo {pages} {56–59}
  (\bibinfo {year} {2020})}\BibitemShut {NoStop}%
\bibitem [{\citenamefont {Bang}\ \emph {et~al.}(2020)\citenamefont {Bang},
  \citenamefont {Seberson}, \citenamefont {Ju}, \citenamefont {Ahn},
  \citenamefont {Xu}, \citenamefont {Gao}, \citenamefont {Robicheaux},\ and\
  \citenamefont {Li}}]{bang2020}%
  \BibitemOpen
  \bibfield  {author} {\bibinfo {author} {\bibfnamefont {J.}~\bibnamefont
  {Bang}}, \bibinfo {author} {\bibfnamefont {T.}~\bibnamefont {Seberson}},
  \bibinfo {author} {\bibfnamefont {P.}~\bibnamefont {Ju}}, \bibinfo {author}
  {\bibfnamefont {J.}~\bibnamefont {Ahn}}, \bibinfo {author} {\bibfnamefont
  {Z.}~\bibnamefont {Xu}}, \bibinfo {author} {\bibfnamefont {X.}~\bibnamefont
  {Gao}}, \bibinfo {author} {\bibfnamefont {F.}~\bibnamefont {Robicheaux}}, \
  and\ \bibinfo {author} {\bibfnamefont {T.}~\bibnamefont {Li}},\ }\bibfield
  {title} {\emph {\bibinfo {title} {Five-dimensional cooling and nonlinear
  dynamics of an optically levitated nanodumbbell},\ }}\href {\doibase
  10.1103/PhysRevResearch.2.043054} {\bibfield  {journal} {\bibinfo  {journal}
  {Phys. Rev. Research}\ }\textbf {\bibinfo {volume} {2}},\ \bibinfo {pages}
  {043054} (\bibinfo {year} {2020})}\BibitemShut {NoStop}%
\bibitem [{\citenamefont {van~der Laan}\ \emph
  {et~al.}(2020{\natexlab{a}})\citenamefont {van~der Laan}, \citenamefont
  {Reimann}, \citenamefont {Vijayan}, \citenamefont {Novotny},\ and\
  \citenamefont {Frimmer}}]{vanderlaan2020b}%
  \BibitemOpen
  \bibfield  {author} {\bibinfo {author} {\bibfnamefont {F.}~\bibnamefont
  {van~der Laan}}, \bibinfo {author} {\bibfnamefont {R.}~\bibnamefont
  {Reimann}}, \bibinfo {author} {\bibfnamefont {F.~T.~J.}\ \bibnamefont
  {Vijayan}}, \bibinfo {author} {\bibfnamefont {L.}~\bibnamefont {Novotny}}, \
  and\ \bibinfo {author} {\bibfnamefont {M.}~\bibnamefont {Frimmer}},\
  }\bibfield  {title} {\emph {\bibinfo {title} {Observation of radiation torque
  shot noise on an optically levitated nanodumbbell},\ }}\href@noop {}
  {\bibfield  {journal} {\bibinfo  {journal} {arXiv preprint arXiv:2012.14231}\
  } (\bibinfo {year} {2020}{\natexlab{a}})}\BibitemShut {NoStop}%
\bibitem [{\citenamefont {Stickler}\ \emph
  {et~al.}(2018{\natexlab{a}})\citenamefont {Stickler}, \citenamefont
  {Papendell}, \citenamefont {Kuhn}, \citenamefont {Schrinski}, \citenamefont
  {Millen}, \citenamefont {Arndt},\ and\ \citenamefont
  {Hornberger}}]{stickler2018b}%
  \BibitemOpen
  \bibfield  {author} {\bibinfo {author} {\bibfnamefont {B.~A.}\ \bibnamefont
  {Stickler}}, \bibinfo {author} {\bibfnamefont {B.}~\bibnamefont {Papendell}},
  \bibinfo {author} {\bibfnamefont {S.}~\bibnamefont {Kuhn}}, \bibinfo {author}
  {\bibfnamefont {B.}~\bibnamefont {Schrinski}}, \bibinfo {author}
  {\bibfnamefont {J.}~\bibnamefont {Millen}}, \bibinfo {author} {\bibfnamefont
  {M.}~\bibnamefont {Arndt}}, \ and\ \bibinfo {author} {\bibfnamefont
  {K.}~\bibnamefont {Hornberger}},\ }\bibfield  {title} {\emph {\bibinfo
  {title} {Probing macroscopic quantum superpositions with nanorotors},\
  }}\href@noop {} {\bibfield  {journal} {\bibinfo  {journal} {New J. Phys.}\
  }\textbf {\bibinfo {volume} {20}},\ \bibinfo {pages} {122001} (\bibinfo
  {year} {2018}{\natexlab{a}})}\BibitemShut {NoStop}%
\bibitem [{\citenamefont {Ma}\ \emph {et~al.}(2020)\citenamefont {Ma},
  \citenamefont {Khosla}, \citenamefont {Stickler},\ and\ \citenamefont
  {Kim}}]{ma2020}%
  \BibitemOpen
  \bibfield  {author} {\bibinfo {author} {\bibfnamefont {Y.}~\bibnamefont
  {Ma}}, \bibinfo {author} {\bibfnamefont {K.~E.}\ \bibnamefont {Khosla}},
  \bibinfo {author} {\bibfnamefont {B.~A.}\ \bibnamefont {Stickler}}, \ and\
  \bibinfo {author} {\bibfnamefont {M.~S.}\ \bibnamefont {Kim}},\ }\bibfield
  {title} {\emph {\bibinfo {title} {Quantum persistent tennis racket dynamics
  of nanorotors},\ }}\href {\doibase 10.1103/PhysRevLett.125.053604} {\bibfield
   {journal} {\bibinfo  {journal} {Phys. Rev. Lett.}\ }\textbf {\bibinfo
  {volume} {125}},\ \bibinfo {pages} {053604} (\bibinfo {year}
  {2020})}\BibitemShut {NoStop}%
\bibitem [{\citenamefont {Rusconi}\ \emph
  {et~al.}(2017{\natexlab{a}})\citenamefont {Rusconi}, \citenamefont
  {P{\"o}chhacker}, \citenamefont {Kustura}, \citenamefont {Cirac},\ and\
  \citenamefont {Romero-Isart}}]{rusconi2017a}%
  \BibitemOpen
  \bibfield  {author} {\bibinfo {author} {\bibfnamefont {C.~C.}\ \bibnamefont
  {Rusconi}}, \bibinfo {author} {\bibfnamefont {V.}~\bibnamefont
  {P{\"o}chhacker}}, \bibinfo {author} {\bibfnamefont {K.}~\bibnamefont
  {Kustura}}, \bibinfo {author} {\bibfnamefont {J.~I.}\ \bibnamefont {Cirac}},
  \ and\ \bibinfo {author} {\bibfnamefont {O.}~\bibnamefont {Romero-Isart}},\
  }\bibfield  {title} {\emph {\bibinfo {title} {Quantum spin stabilized
  magnetic levitation},\ }}\href@noop {} {\bibfield  {journal} {\bibinfo
  {journal} {Phys. Rev. Lett.}\ }\textbf {\bibinfo {volume} {119}},\ \bibinfo
  {pages} {167202} (\bibinfo {year} {2017}{\natexlab{a}})}\BibitemShut
  {NoStop}%
\bibitem [{\citenamefont {Schrinski}\ \emph {et~al.}(2019)\citenamefont
  {Schrinski}, \citenamefont {Nimmrichter}, \citenamefont {Stickler},\ and\
  \citenamefont {Hornberger}}]{schrinski2019}%
  \BibitemOpen
  \bibfield  {author} {\bibinfo {author} {\bibfnamefont {B.}~\bibnamefont
  {Schrinski}}, \bibinfo {author} {\bibfnamefont {S.}~\bibnamefont
  {Nimmrichter}}, \bibinfo {author} {\bibfnamefont {B.~A.}\ \bibnamefont
  {Stickler}}, \ and\ \bibinfo {author} {\bibfnamefont {K.}~\bibnamefont
  {Hornberger}},\ }\bibfield  {title} {\emph {\bibinfo {title} {Macroscopicity
  of quantum mechanical superposition tests via hypothesis falsification},\
  }}\href {\doibase 10.1103/PhysRevA.100.032111} {\bibfield  {journal}
  {\bibinfo  {journal} {Phys. Rev. A}\ }\textbf {\bibinfo {volume} {100}},\
  \bibinfo {pages} {032111} (\bibinfo {year} {2019})}\BibitemShut {NoStop}%
\bibitem [{\citenamefont {Gieseler}\ \emph
  {et~al.}(2020{\natexlab{a}})\citenamefont {Gieseler}, \citenamefont
  {Gomez-Solano}, \citenamefont {Magazz{\`u}}, \citenamefont {Castillo},
  \citenamefont {Garc{\'\i}a}, \citenamefont {Gironella-Torrent}, \citenamefont
  {Viader-Godoy}, \citenamefont {Ritort}, \citenamefont {Pesce}, \citenamefont
  {Arzola} \emph {et~al.}}]{gieseler2020}%
  \BibitemOpen
  \bibfield  {author} {\bibinfo {author} {\bibfnamefont {J.}~\bibnamefont
  {Gieseler}}, \bibinfo {author} {\bibfnamefont {J.~R.}\ \bibnamefont
  {Gomez-Solano}}, \bibinfo {author} {\bibfnamefont {A.}~\bibnamefont
  {Magazz{\`u}}}, \bibinfo {author} {\bibfnamefont {I.~P.}\ \bibnamefont
  {Castillo}}, \bibinfo {author} {\bibfnamefont {L.~P.}\ \bibnamefont
  {Garc{\'\i}a}}, \bibinfo {author} {\bibfnamefont {M.}~\bibnamefont
  {Gironella-Torrent}}, \bibinfo {author} {\bibfnamefont {X.}~\bibnamefont
  {Viader-Godoy}}, \bibinfo {author} {\bibfnamefont {F.}~\bibnamefont
  {Ritort}}, \bibinfo {author} {\bibfnamefont {G.}~\bibnamefont {Pesce}},
  \bibinfo {author} {\bibfnamefont {A.~V.}\ \bibnamefont {Arzola}},  \emph
  {et~al.},\ }\bibfield  {title} {\emph {\bibinfo {title} {Optical tweezers: A
  comprehensive tutorial from calibration to applications},\ }}\href@noop {}
  {\bibfield  {journal} {\bibinfo  {journal} {arXiv preprint arXiv:2004.05246}\
  } (\bibinfo {year} {2020}{\natexlab{a}})}\BibitemShut {NoStop}%
\bibitem [{\citenamefont {Kuhn}\ \emph {et~al.}(2015)\citenamefont {Kuhn},
  \citenamefont {Asenbaum}, \citenamefont {Kosloff}, \citenamefont {Sclafani},
  \citenamefont {Stickler}, \citenamefont {Nimmrichter}, \citenamefont
  {Hornberger}, \citenamefont {Cheshnovsky}, \citenamefont {Patolsky},\ and\
  \citenamefont {Arndt}}]{kuhn2015}%
  \BibitemOpen
  \bibfield  {author} {\bibinfo {author} {\bibfnamefont {S.}~\bibnamefont
  {Kuhn}}, \bibinfo {author} {\bibfnamefont {P.}~\bibnamefont {Asenbaum}},
  \bibinfo {author} {\bibfnamefont {A.}~\bibnamefont {Kosloff}}, \bibinfo
  {author} {\bibfnamefont {M.}~\bibnamefont {Sclafani}}, \bibinfo {author}
  {\bibfnamefont {B.~A.}\ \bibnamefont {Stickler}}, \bibinfo {author}
  {\bibfnamefont {S.}~\bibnamefont {Nimmrichter}}, \bibinfo {author}
  {\bibfnamefont {K.}~\bibnamefont {Hornberger}}, \bibinfo {author}
  {\bibfnamefont {O.}~\bibnamefont {Cheshnovsky}}, \bibinfo {author}
  {\bibfnamefont {F.}~\bibnamefont {Patolsky}}, \ and\ \bibinfo {author}
  {\bibfnamefont {M.}~\bibnamefont {Arndt}},\ }\bibfield  {title} {\emph
  {\bibinfo {title} {{Cavity-Assisted Manipulation of Freely Rotating Silicon
  Nanorods in High Vacuum}},\ }}\href@noop {} {\bibfield  {journal} {\bibinfo
  {journal} {Nano Lett.}\ }\textbf {\bibinfo {volume} {15}},\ \bibinfo {pages}
  {5604} (\bibinfo {year} {2015})}\BibitemShut {NoStop}%
\bibitem [{\citenamefont {Stickler}\ \emph
  {et~al.}(2016{\natexlab{a}})\citenamefont {Stickler}, \citenamefont
  {Nimmrichter}, \citenamefont {Martinetz}, \citenamefont {Kuhn}, \citenamefont
  {Arndt},\ and\ \citenamefont {Hornberger}}]{stickler2016a}%
  \BibitemOpen
  \bibfield  {author} {\bibinfo {author} {\bibfnamefont {B.~A.}\ \bibnamefont
  {Stickler}}, \bibinfo {author} {\bibfnamefont {S.}~\bibnamefont
  {Nimmrichter}}, \bibinfo {author} {\bibfnamefont {L.}~\bibnamefont
  {Martinetz}}, \bibinfo {author} {\bibfnamefont {S.}~\bibnamefont {Kuhn}},
  \bibinfo {author} {\bibfnamefont {M.}~\bibnamefont {Arndt}}, \ and\ \bibinfo
  {author} {\bibfnamefont {K.}~\bibnamefont {Hornberger}},\ }\bibfield  {title}
  {\emph {\bibinfo {title} {Rotranslational cavity cooling of dielectric rods
  and disks},\ }}\href@noop {} {\bibfield  {journal} {\bibinfo  {journal}
  {Phys. Rev. A}\ }\textbf {\bibinfo {volume} {94}},\ \bibinfo {pages} {033818}
  (\bibinfo {year} {2016}{\natexlab{a}})}\BibitemShut {NoStop}%
\bibitem [{\citenamefont {van~der Laan}\ \emph
  {et~al.}(2020{\natexlab{b}})\citenamefont {van~der Laan}, \citenamefont
  {Reimann}, \citenamefont {Militaru}, \citenamefont {Tebbenjohanns},
  \citenamefont {Windey}, \citenamefont {Frimmer},\ and\ \citenamefont
  {Novotny}}]{vanderlaan2020}%
  \BibitemOpen
  \bibfield  {author} {\bibinfo {author} {\bibfnamefont {F.}~\bibnamefont
  {van~der Laan}}, \bibinfo {author} {\bibfnamefont {R.}~\bibnamefont
  {Reimann}}, \bibinfo {author} {\bibfnamefont {A.}~\bibnamefont {Militaru}},
  \bibinfo {author} {\bibfnamefont {F.}~\bibnamefont {Tebbenjohanns}}, \bibinfo
  {author} {\bibfnamefont {D.}~\bibnamefont {Windey}}, \bibinfo {author}
  {\bibfnamefont {M.}~\bibnamefont {Frimmer}}, \ and\ \bibinfo {author}
  {\bibfnamefont {L.}~\bibnamefont {Novotny}},\ }\bibfield  {title} {\emph
  {\bibinfo {title} {Optically levitated rotor at its thermal limit of
  frequency stability},\ }}\href {\doibase 10.1103/PhysRevA.102.013505}
  {\bibfield  {journal} {\bibinfo  {journal} {Phys. Rev. A}\ }\textbf {\bibinfo
  {volume} {102}},\ \bibinfo {pages} {013505} (\bibinfo {year}
  {2020}{\natexlab{b}})}\BibitemShut {NoStop}%
\bibitem [{\citenamefont {Zhong}\ and\ \citenamefont
  {Robicheaux}(2017)}]{zhong2017}%
  \BibitemOpen
  \bibfield  {author} {\bibinfo {author} {\bibfnamefont {C.}~\bibnamefont
  {Zhong}}\ and\ \bibinfo {author} {\bibfnamefont {F.}~\bibnamefont
  {Robicheaux}},\ }\bibfield  {title} {\emph {\bibinfo {title}
  {Shot-noise-dominant regime for ellipsoidal nanoparticles in a linearly
  polarized beam},\ }}\href@noop {} {\bibfield  {journal} {\bibinfo  {journal}
  {Phys. Rev. A}\ }\textbf {\bibinfo {volume} {95}},\ \bibinfo {pages} {053421}
  (\bibinfo {year} {2017})}\BibitemShut {NoStop}%
\bibitem [{\citenamefont {Seberson}\ and\ \citenamefont
  {Robicheaux}(2019)}]{seberson2019}%
  \BibitemOpen
  \bibfield  {author} {\bibinfo {author} {\bibfnamefont {T.}~\bibnamefont
  {Seberson}}\ and\ \bibinfo {author} {\bibfnamefont {F.}~\bibnamefont
  {Robicheaux}},\ }\bibfield  {title} {\emph {\bibinfo {title} {Parametric
  feedback cooling of rigid body nanodumbbells in levitated optomechanics},\
  }}\href {\doibase 10.1103/PhysRevA.99.013821} {\bibfield  {journal} {\bibinfo
   {journal} {Phys. Rev. A}\ }\textbf {\bibinfo {volume} {99}},\ \bibinfo
  {pages} {013821} (\bibinfo {year} {2019})}\BibitemShut {NoStop}%
\bibitem [{\citenamefont {Sch{\"a}fer}\ \emph {et~al.}(2020)\citenamefont
  {Sch{\"a}fer}, \citenamefont {Rudolph}, \citenamefont {Hornberger},\ and\
  \citenamefont {Stickler}}]{schafer2020}%
  \BibitemOpen
  \bibfield  {author} {\bibinfo {author} {\bibfnamefont {J.}~\bibnamefont
  {Sch{\"a}fer}}, \bibinfo {author} {\bibfnamefont {H.}~\bibnamefont
  {Rudolph}}, \bibinfo {author} {\bibfnamefont {K.}~\bibnamefont {Hornberger}},
  \ and\ \bibinfo {author} {\bibfnamefont {B.~A.}\ \bibnamefont {Stickler}},\
  }\bibfield  {title} {\emph {\bibinfo {title} {Cooling nanorotors by elliptic
  coherent scattering},\ }}\href@noop {} {\bibfield  {journal} {\bibinfo
  {journal} {arXiv preprint arXiv:2006.04090}\ } (\bibinfo {year}
  {2020})}\BibitemShut {NoStop}%
\bibitem [{\citenamefont {Arita}\ \emph {et~al.}(2013)\citenamefont {Arita},
  \citenamefont {Mazilu},\ and\ \citenamefont {Dholakia}}]{arita2013}%
  \BibitemOpen
  \bibfield  {author} {\bibinfo {author} {\bibfnamefont {Y.}~\bibnamefont
  {Arita}}, \bibinfo {author} {\bibfnamefont {M.}~\bibnamefont {Mazilu}}, \
  and\ \bibinfo {author} {\bibfnamefont {K.}~\bibnamefont {Dholakia}},\
  }\bibfield  {title} {\emph {\bibinfo {title} {Laser-induced rotation and
  cooling of a trapped microgyroscope in vacuum},\ }}\href@noop {} {\bibfield
  {journal} {\bibinfo  {journal} {Nat. Commun.}\ }\textbf {\bibinfo {volume}
  {4}},\ \bibinfo {pages} {2374} (\bibinfo {year} {2013})}\BibitemShut
  {NoStop}%
\bibitem [{\citenamefont {Monteiro}\ \emph {et~al.}(2018)\citenamefont
  {Monteiro}, \citenamefont {Ghosh}, \citenamefont {van Assendelft},\ and\
  \citenamefont {Moore}}]{monteiro2018}%
  \BibitemOpen
  \bibfield  {author} {\bibinfo {author} {\bibfnamefont {F.}~\bibnamefont
  {Monteiro}}, \bibinfo {author} {\bibfnamefont {S.}~\bibnamefont {Ghosh}},
  \bibinfo {author} {\bibfnamefont {E.~C.}\ \bibnamefont {van Assendelft}}, \
  and\ \bibinfo {author} {\bibfnamefont {D.~C.}\ \bibnamefont {Moore}},\
  }\bibfield  {title} {\emph {\bibinfo {title} {Optical rotation of levitated
  spheres in high vacuum},\ }}\href@noop {} {\bibfield  {journal} {\bibinfo
  {journal} {Phys. Rev. A}\ }\textbf {\bibinfo {volume} {97}},\ \bibinfo
  {pages} {051802} (\bibinfo {year} {2018})}\BibitemShut {NoStop}%
\bibitem [{\citenamefont {Martinetz}\ \emph {et~al.}(2018)\citenamefont
  {Martinetz}, \citenamefont {Hornberger},\ and\ \citenamefont
  {Stickler}}]{martinetz2018}%
  \BibitemOpen
  \bibfield  {author} {\bibinfo {author} {\bibfnamefont {L.}~\bibnamefont
  {Martinetz}}, \bibinfo {author} {\bibfnamefont {K.}~\bibnamefont
  {Hornberger}}, \ and\ \bibinfo {author} {\bibfnamefont {B.~A.}\ \bibnamefont
  {Stickler}},\ }\bibfield  {title} {\emph {\bibinfo {title} {Gas-induced
  friction and diffusion of rigid rotors},\ }}\href@noop {} {\bibfield
  {journal} {\bibinfo  {journal} {Phys. Rev. E}\ }\textbf {\bibinfo {volume}
  {97}},\ \bibinfo {pages} {052112} (\bibinfo {year} {2018})}\BibitemShut
  {NoStop}%
\bibitem [{\citenamefont {H\"ummer}\ \emph {et~al.}(2020)\citenamefont
  {H\"ummer}, \citenamefont {Lampert}, \citenamefont {Kustura}, \citenamefont
  {Maurer}, \citenamefont {Gonzalez-Ballestero},\ and\ \citenamefont
  {Romero-Isart}}]{hummer2020}%
  \BibitemOpen
  \bibfield  {author} {\bibinfo {author} {\bibfnamefont {D.}~\bibnamefont
  {H\"ummer}}, \bibinfo {author} {\bibfnamefont {R.}~\bibnamefont {Lampert}},
  \bibinfo {author} {\bibfnamefont {K.}~\bibnamefont {Kustura}}, \bibinfo
  {author} {\bibfnamefont {P.}~\bibnamefont {Maurer}}, \bibinfo {author}
  {\bibfnamefont {C.}~\bibnamefont {Gonzalez-Ballestero}}, \ and\ \bibinfo
  {author} {\bibfnamefont {O.}~\bibnamefont {Romero-Isart}},\ }\bibfield
  {title} {\emph {\bibinfo {title} {Acoustic and optical properties of a
  fast-spinning dielectric nanoparticle},\ }}\href {\doibase
  10.1103/PhysRevB.101.205416} {\bibfield  {journal} {\bibinfo  {journal}
  {Phys. Rev. B}\ }\textbf {\bibinfo {volume} {101}},\ \bibinfo {pages}
  {205416} (\bibinfo {year} {2020})}\BibitemShut {NoStop}%
\bibitem [{\citenamefont {Arita}\ \emph {et~al.}(2020)\citenamefont {Arita},
  \citenamefont {Simpson}, \citenamefont {Zem{\'a}nek},\ and\ \citenamefont
  {Dholakia}}]{arita2020}%
  \BibitemOpen
  \bibfield  {author} {\bibinfo {author} {\bibfnamefont {Y.}~\bibnamefont
  {Arita}}, \bibinfo {author} {\bibfnamefont {S.~H.}\ \bibnamefont {Simpson}},
  \bibinfo {author} {\bibfnamefont {P.}~\bibnamefont {Zem{\'a}nek}}, \ and\
  \bibinfo {author} {\bibfnamefont {K.}~\bibnamefont {Dholakia}},\ }\bibfield
  {title} {\emph {\bibinfo {title} {Coherent oscillations of a levitated
  birefringent microsphere in vacuum driven by nonconservative
  rotation-translation coupling},\ }}\href@noop {} {\bibfield  {journal}
  {\bibinfo  {journal} {Sci. Adv.}\ }\textbf {\bibinfo {volume} {6}} (\bibinfo
  {year} {2020})}\BibitemShut {NoStop}%
\bibitem [{\citenamefont {Kane}(2010)}]{kane2010}%
  \BibitemOpen
  \bibfield  {author} {\bibinfo {author} {\bibfnamefont {B.}~\bibnamefont
  {Kane}},\ }\bibfield  {title} {\emph {\bibinfo {title} {Levitated spinning
  graphene flakes in an electric quadrupole ion trap},\ }}\href@noop {}
  {\bibfield  {journal} {\bibinfo  {journal} {Phys. Rev. B}\ }\textbf {\bibinfo
  {volume} {82}},\ \bibinfo {pages} {115441} (\bibinfo {year}
  {2010})}\BibitemShut {NoStop}%
\bibitem [{\citenamefont {Millen}\ \emph {et~al.}(2015)\citenamefont {Millen},
  \citenamefont {Fonseca}, \citenamefont {Mavrogordatos}, \citenamefont
  {Monteiro},\ and\ \citenamefont {Barker}}]{millen2015}%
  \BibitemOpen
  \bibfield  {author} {\bibinfo {author} {\bibfnamefont {J.}~\bibnamefont
  {Millen}}, \bibinfo {author} {\bibfnamefont {P.}~\bibnamefont {Fonseca}},
  \bibinfo {author} {\bibfnamefont {T.}~\bibnamefont {Mavrogordatos}}, \bibinfo
  {author} {\bibfnamefont {T.}~\bibnamefont {Monteiro}}, \ and\ \bibinfo
  {author} {\bibfnamefont {P.}~\bibnamefont {Barker}},\ }\bibfield  {title}
  {\emph {\bibinfo {title} {Cavity cooling a single charged levitated
  nanosphere},\ }}\href@noop {} {\bibfield  {journal} {\bibinfo  {journal}
  {Phys. Rev. Lett.}\ }\textbf {\bibinfo {volume} {114}},\ \bibinfo {pages}
  {123602} (\bibinfo {year} {2015})}\BibitemShut {NoStop}%
\bibitem [{\citenamefont {Delord}\ \emph
  {et~al.}(2017{\natexlab{a}})\citenamefont {Delord}, \citenamefont {Nicolas},
  \citenamefont {Schwab},\ and\ \citenamefont {H{\'e}tet}}]{delord2017b}%
  \BibitemOpen
  \bibfield  {author} {\bibinfo {author} {\bibfnamefont {T.}~\bibnamefont
  {Delord}}, \bibinfo {author} {\bibfnamefont {L.}~\bibnamefont {Nicolas}},
  \bibinfo {author} {\bibfnamefont {L.}~\bibnamefont {Schwab}}, \ and\ \bibinfo
  {author} {\bibfnamefont {G.}~\bibnamefont {H{\'e}tet}},\ }\bibfield  {title}
  {\emph {\bibinfo {title} {Electron spin resonance from nv centers in diamonds
  levitating in an ion trap},\ }}\href@noop {} {\bibfield  {journal} {\bibinfo
  {journal} {New J. Phys.}\ }\textbf {\bibinfo {volume} {19}},\ \bibinfo
  {pages} {033031} (\bibinfo {year} {2017}{\natexlab{a}})}\BibitemShut
  {NoStop}%
\bibitem [{\citenamefont {Nagornykh}\ \emph {et~al.}(2017)\citenamefont
  {Nagornykh}, \citenamefont {Coppock}, \citenamefont {Murphy},\ and\
  \citenamefont {Kane}}]{nagornykh2017}%
  \BibitemOpen
  \bibfield  {author} {\bibinfo {author} {\bibfnamefont {P.}~\bibnamefont
  {Nagornykh}}, \bibinfo {author} {\bibfnamefont {J.~E.}\ \bibnamefont
  {Coppock}}, \bibinfo {author} {\bibfnamefont {J.~P.}\ \bibnamefont {Murphy}},
  \ and\ \bibinfo {author} {\bibfnamefont {B.}~\bibnamefont {Kane}},\
  }\bibfield  {title} {\emph {\bibinfo {title} {Optical and magnetic
  measurements of gyroscopically stabilized graphene nanoplatelets levitated in
  an ion trap},\ }}\href@noop {} {\bibfield  {journal} {\bibinfo  {journal}
  {Phys. Rev. B}\ }\textbf {\bibinfo {volume} {96}},\ \bibinfo {pages} {035402}
  (\bibinfo {year} {2017})}\BibitemShut {NoStop}%
\bibitem [{\citenamefont {Bykov}\ \emph {et~al.}(2019)\citenamefont {Bykov},
  \citenamefont {Mestres}, \citenamefont {Dania}, \citenamefont
  {Schm{\"o}ger},\ and\ \citenamefont {Northup}}]{bykov2019}%
  \BibitemOpen
  \bibfield  {author} {\bibinfo {author} {\bibfnamefont {D.~S.}\ \bibnamefont
  {Bykov}}, \bibinfo {author} {\bibfnamefont {P.}~\bibnamefont {Mestres}},
  \bibinfo {author} {\bibfnamefont {L.}~\bibnamefont {Dania}}, \bibinfo
  {author} {\bibfnamefont {L.}~\bibnamefont {Schm{\"o}ger}}, \ and\ \bibinfo
  {author} {\bibfnamefont {T.~E.}\ \bibnamefont {Northup}},\ }\bibfield
  {title} {\emph {\bibinfo {title} {Direct loading of nanoparticles under high
  vacuum into a paul trap for levitodynamical experiments},\ }}\href@noop {}
  {\bibfield  {journal} {\bibinfo  {journal} {Appl. Phys. Lett.}\ }\textbf
  {\bibinfo {volume} {115}},\ \bibinfo {pages} {034101} (\bibinfo {year}
  {2019})}\BibitemShut {NoStop}%
\bibitem [{\citenamefont {Moore}\ \emph {et~al.}(2014)\citenamefont {Moore},
  \citenamefont {Rider},\ and\ \citenamefont {Gratta}}]{moore2014}%
  \BibitemOpen
  \bibfield  {author} {\bibinfo {author} {\bibfnamefont {D.~C.}\ \bibnamefont
  {Moore}}, \bibinfo {author} {\bibfnamefont {A.~D.}\ \bibnamefont {Rider}}, \
  and\ \bibinfo {author} {\bibfnamefont {G.}~\bibnamefont {Gratta}},\
  }\bibfield  {title} {\emph {\bibinfo {title} {Search for millicharged
  particles using optically levitated microspheres},\ }}\href {\doibase
  10.1103/PhysRevLett.113.251801} {\bibfield  {journal} {\bibinfo  {journal}
  {Phys. Rev. Lett.}\ }\textbf {\bibinfo {volume} {113}},\ \bibinfo {pages}
  {251801} (\bibinfo {year} {2014})}\BibitemShut {NoStop}%
\bibitem [{\citenamefont {Frimmer}\ \emph {et~al.}(2017)\citenamefont
  {Frimmer}, \citenamefont {Luszcz}, \citenamefont {Ferreiro}, \citenamefont
  {Jain}, \citenamefont {Hebestreit},\ and\ \citenamefont
  {Novotny}}]{frimmer2017}%
  \BibitemOpen
  \bibfield  {author} {\bibinfo {author} {\bibfnamefont {M.}~\bibnamefont
  {Frimmer}}, \bibinfo {author} {\bibfnamefont {K.}~\bibnamefont {Luszcz}},
  \bibinfo {author} {\bibfnamefont {S.}~\bibnamefont {Ferreiro}}, \bibinfo
  {author} {\bibfnamefont {V.}~\bibnamefont {Jain}}, \bibinfo {author}
  {\bibfnamefont {E.}~\bibnamefont {Hebestreit}}, \ and\ \bibinfo {author}
  {\bibfnamefont {L.}~\bibnamefont {Novotny}},\ }\bibfield  {title} {\emph
  {\bibinfo {title} {Controlling the net charge on a nanoparticle optically
  levitated in vacuum},\ }}\href@noop {} {\bibfield  {journal} {\bibinfo
  {journal} {Phys. Rev. A}\ }\textbf {\bibinfo {volume} {95}},\ \bibinfo
  {pages} {061801} (\bibinfo {year} {2017})}\BibitemShut {NoStop}%
\bibitem [{\citenamefont {Goldwater}\ \emph {et~al.}(2019)\citenamefont
  {Goldwater}, \citenamefont {Stickler}, \citenamefont {Martinetz},
  \citenamefont {Northup}, \citenamefont {Hornberger},\ and\ \citenamefont
  {Millen}}]{goldwater2019}%
  \BibitemOpen
  \bibfield  {author} {\bibinfo {author} {\bibfnamefont {D.}~\bibnamefont
  {Goldwater}}, \bibinfo {author} {\bibfnamefont {B.~A.}\ \bibnamefont
  {Stickler}}, \bibinfo {author} {\bibfnamefont {L.}~\bibnamefont {Martinetz}},
  \bibinfo {author} {\bibfnamefont {T.~E.}\ \bibnamefont {Northup}}, \bibinfo
  {author} {\bibfnamefont {K.}~\bibnamefont {Hornberger}}, \ and\ \bibinfo
  {author} {\bibfnamefont {J.}~\bibnamefont {Millen}},\ }\bibfield  {title}
  {\emph {\bibinfo {title} {Levitated electromechanics: all-electrical cooling
  of charged nano- and micro-particles},\ }}\href@noop {} {\bibfield  {journal}
  {\bibinfo  {journal} {Quant. Sci. Techn.}\ }\textbf {\bibinfo {volume} {4}},\
  \bibinfo {pages} {024003} (\bibinfo {year} {2019})}\BibitemShut {NoStop}%
\bibitem [{\citenamefont {Martinetz}\ \emph {et~al.}(2020)\citenamefont
  {Martinetz}, \citenamefont {Hornberger}, \citenamefont {Millen},
  \citenamefont {Kim},\ and\ \citenamefont {Stickler}}]{martinetz2020}%
  \BibitemOpen
  \bibfield  {author} {\bibinfo {author} {\bibfnamefont {L.}~\bibnamefont
  {Martinetz}}, \bibinfo {author} {\bibfnamefont {K.}~\bibnamefont
  {Hornberger}}, \bibinfo {author} {\bibfnamefont {J.}~\bibnamefont {Millen}},
  \bibinfo {author} {\bibfnamefont {M.}~\bibnamefont {Kim}}, \ and\ \bibinfo
  {author} {\bibfnamefont {B.~A.}\ \bibnamefont {Stickler}},\ }\bibfield
  {title} {\emph {\bibinfo {title} {Quantum electromechanics with levitated
  nanoparticles},\ }}\href@noop {} {\bibfield  {journal} {\bibinfo  {journal}
  {npj Quant. Inf.}\ }\textbf {\bibinfo {volume} {6}} (\bibinfo {year}
  {2020})}\BibitemShut {NoStop}%
\bibitem [{\citenamefont {Delord}\ \emph
  {et~al.}(2017{\natexlab{b}})\citenamefont {Delord}, \citenamefont {Nicolas},
  \citenamefont {Chassagneux},\ and\ \citenamefont {H\'etet}}]{delord2017a}%
  \BibitemOpen
  \bibfield  {author} {\bibinfo {author} {\bibfnamefont {T.}~\bibnamefont
  {Delord}}, \bibinfo {author} {\bibfnamefont {L.}~\bibnamefont {Nicolas}},
  \bibinfo {author} {\bibfnamefont {Y.}~\bibnamefont {Chassagneux}}, \ and\
  \bibinfo {author} {\bibfnamefont {G.}~\bibnamefont {H\'etet}},\ }\bibfield
  {title} {\emph {\bibinfo {title} {Strong coupling between a single
  nitrogen-vacancy spin and the rotational mode of diamonds levitating in an
  ion trap},\ }}\href {\doibase 10.1103/PhysRevA.96.063810} {\bibfield
  {journal} {\bibinfo  {journal} {Phys. Rev. A}\ }\textbf {\bibinfo {volume}
  {96}},\ \bibinfo {pages} {063810} (\bibinfo {year}
  {2017}{\natexlab{b}})}\BibitemShut {NoStop}%
\bibitem [{\citenamefont {Coppock}\ \emph {et~al.}(2016)\citenamefont
  {Coppock}, \citenamefont {Nagornykh}, \citenamefont {Murphy},\ and\
  \citenamefont {Kane}}]{coppock2016}%
  \BibitemOpen
  \bibfield  {author} {\bibinfo {author} {\bibfnamefont {J.~E.}\ \bibnamefont
  {Coppock}}, \bibinfo {author} {\bibfnamefont {P.}~\bibnamefont {Nagornykh}},
  \bibinfo {author} {\bibfnamefont {J.~P.}\ \bibnamefont {Murphy}}, \ and\
  \bibinfo {author} {\bibfnamefont {B.~E.}\ \bibnamefont {Kane}},\ }in\ \href
  {\doibase 10.1117/12.2258066} {\emph {\bibinfo {booktitle} {Optical Trapping
  and Optical Micromanipulation XIII}}},\ \bibinfo {series} {Proceedings of
  SPIE}, Vol.\ \bibinfo {volume} {9922},\ \bibinfo {editor} {edited by\
  \bibinfo {editor} {\bibfnamefont {K.}~\bibnamefont {Dholakia}}\ and\ \bibinfo
  {editor} {\bibfnamefont {G.~C.}\ \bibnamefont {Spalding}}}\ (\bibinfo
  {publisher} {International Society for Optics and Photonics},\ \bibinfo
  {address} {Bellingham, WA},\ \bibinfo {year} {2016})\ p.\ \bibinfo {pages}
  {99220E}\BibitemShut {NoStop}%
\bibitem [{\citenamefont {Cirio}\ \emph {et~al.}(2012)\citenamefont {Cirio},
  \citenamefont {Brennen},\ and\ \citenamefont {Twamley}}]{cirio2012}%
  \BibitemOpen
  \bibfield  {author} {\bibinfo {author} {\bibfnamefont {M.}~\bibnamefont
  {Cirio}}, \bibinfo {author} {\bibfnamefont {G.~K.}\ \bibnamefont {Brennen}},
  \ and\ \bibinfo {author} {\bibfnamefont {J.}~\bibnamefont {Twamley}},\
  }\bibfield  {title} {\emph {\bibinfo {title} {Quantum magnetomechanics:
  Ultrahigh-{$Q$}-levitated mechanical oscillators},\ }}\href {\doibase
  10.1103/PhysRevLett.109.147206} {\bibfield  {journal} {\bibinfo  {journal}
  {Phys. Rev. Lett.}\ }\textbf {\bibinfo {volume} {109}},\ \bibinfo {pages}
  {147206} (\bibinfo {year} {2012})}\BibitemShut {NoStop}%
\bibitem [{\citenamefont {Romero-Isart}\ \emph {et~al.}(2012)\citenamefont
  {Romero-Isart}, \citenamefont {Clemente}, \citenamefont {Navau},
  \citenamefont {Sanchez},\ and\ \citenamefont {Cirac}}]{romeroisart2012}%
  \BibitemOpen
  \bibfield  {author} {\bibinfo {author} {\bibfnamefont {O.}~\bibnamefont
  {Romero-Isart}}, \bibinfo {author} {\bibfnamefont {L.}~\bibnamefont
  {Clemente}}, \bibinfo {author} {\bibfnamefont {C.}~\bibnamefont {Navau}},
  \bibinfo {author} {\bibfnamefont {A.}~\bibnamefont {Sanchez}}, \ and\
  \bibinfo {author} {\bibfnamefont {J.}~\bibnamefont {Cirac}},\ }\bibfield
  {title} {\emph {\bibinfo {title} {Quantum magnetomechanics with levitating
  superconducting microspheres},\ }}\href@noop {} {\bibfield  {journal}
  {\bibinfo  {journal} {Phys. Rev. Lett.}\ }\textbf {\bibinfo {volume} {109}},\
  \bibinfo {pages} {147205} (\bibinfo {year} {2012})}\BibitemShut {NoStop}%
\bibitem [{\citenamefont {Pino}\ \emph {et~al.}(2018)\citenamefont {Pino},
  \citenamefont {Prat-Camps}, \citenamefont {Sinha}, \citenamefont
  {Venkatesh},\ and\ \citenamefont {Romero-Isart}}]{pino2018}%
  \BibitemOpen
  \bibfield  {author} {\bibinfo {author} {\bibfnamefont {H.}~\bibnamefont
  {Pino}}, \bibinfo {author} {\bibfnamefont {J.}~\bibnamefont {Prat-Camps}},
  \bibinfo {author} {\bibfnamefont {K.}~\bibnamefont {Sinha}}, \bibinfo
  {author} {\bibfnamefont {B.~P.}\ \bibnamefont {Venkatesh}}, \ and\ \bibinfo
  {author} {\bibfnamefont {O.}~\bibnamefont {Romero-Isart}},\ }\bibfield
  {title} {\emph {\bibinfo {title} {On-chip quantum interference of a
  superconducting microsphere},\ }}\href@noop {} {\bibfield  {journal}
  {\bibinfo  {journal} {Quant. Sci. Techn.}\ }\textbf {\bibinfo {volume} {3}},\
  \bibinfo {pages} {025001} (\bibinfo {year} {2018})}\BibitemShut {NoStop}%
\bibitem [{\citenamefont {Einstein}\ and\ \citenamefont
  {De~Haas}(1915)}]{einstein1915}%
  \BibitemOpen
  \bibfield  {author} {\bibinfo {author} {\bibfnamefont {A.}~\bibnamefont
  {Einstein}}\ and\ \bibinfo {author} {\bibfnamefont {W.}~\bibnamefont
  {De~Haas}},\ }in\ \href@noop {} {\emph {\bibinfo {booktitle} {Proc. KNAW}}},\
  Vol.~\bibinfo {volume} {18}\ (\bibinfo {year} {1915})\ p.\ \bibinfo {pages}
  {696}\BibitemShut {NoStop}%
\bibitem [{\citenamefont {Barnett}(1915)}]{barnett1915}%
  \BibitemOpen
  \bibfield  {author} {\bibinfo {author} {\bibfnamefont {S.~J.}\ \bibnamefont
  {Barnett}},\ }\bibfield  {title} {\emph {\bibinfo {title} {Magnetization by
  rotation},\ }}\href {\doibase 10.1103/PhysRev.6.239} {\bibfield  {journal}
  {\bibinfo  {journal} {Phys. Rev.}\ }\textbf {\bibinfo {volume} {6}},\
  \bibinfo {pages} {239} (\bibinfo {year} {1915})}\BibitemShut {NoStop}%
\bibitem [{\citenamefont {Hsu}\ \emph {et~al.}(2016)\citenamefont {Hsu},
  \citenamefont {Ji}, \citenamefont {Lewandowski},\ and\ \citenamefont
  {D’Urso}}]{hsu2016}%
  \BibitemOpen
  \bibfield  {author} {\bibinfo {author} {\bibfnamefont {J.-F.}\ \bibnamefont
  {Hsu}}, \bibinfo {author} {\bibfnamefont {P.}~\bibnamefont {Ji}}, \bibinfo
  {author} {\bibfnamefont {C.~W.}\ \bibnamefont {Lewandowski}}, \ and\ \bibinfo
  {author} {\bibfnamefont {B.}~\bibnamefont {D’Urso}},\ }\bibfield  {title}
  {\emph {\bibinfo {title} {Cooling the motion of diamond nanocrystals in a
  magneto-gravitational trap in high vacuum},\ }}\href@noop {} {\bibfield
  {journal} {\bibinfo  {journal} {Sci. Rep.}\ }\textbf {\bibinfo {volume}
  {6}},\ \bibinfo {pages} {30125} (\bibinfo {year} {2016})}\BibitemShut
  {NoStop}%
\bibitem [{\citenamefont {Slezak}\ \emph {et~al.}(2018)\citenamefont {Slezak},
  \citenamefont {Lewandowski}, \citenamefont {Hsu},\ and\ \citenamefont
  {D’Urso}}]{slezak2018}%
  \BibitemOpen
  \bibfield  {author} {\bibinfo {author} {\bibfnamefont {B.~R.}\ \bibnamefont
  {Slezak}}, \bibinfo {author} {\bibfnamefont {C.~W.}\ \bibnamefont
  {Lewandowski}}, \bibinfo {author} {\bibfnamefont {J.-F.}\ \bibnamefont
  {Hsu}}, \ and\ \bibinfo {author} {\bibfnamefont {B.}~\bibnamefont
  {D’Urso}},\ }\bibfield  {title} {\emph {\bibinfo {title} {Cooling the
  motion of a silica microsphere in a magneto-gravitational trap in ultra-high
  vacuum},\ }}\href@noop {} {\bibfield  {journal} {\bibinfo  {journal} {New J.
  Phys.}\ }\textbf {\bibinfo {volume} {20}},\ \bibinfo {pages} {063028}
  (\bibinfo {year} {2018})}\BibitemShut {NoStop}%
\bibitem [{\citenamefont {O'Brien}\ \emph {et~al.}(2019)\citenamefont
  {O'Brien}, \citenamefont {Dunn}, \citenamefont {Downes},\ and\ \citenamefont
  {Twamley}}]{obrian2019}%
  \BibitemOpen
  \bibfield  {author} {\bibinfo {author} {\bibfnamefont {M.}~\bibnamefont
  {O'Brien}}, \bibinfo {author} {\bibfnamefont {S.}~\bibnamefont {Dunn}},
  \bibinfo {author} {\bibfnamefont {J.}~\bibnamefont {Downes}}, \ and\ \bibinfo
  {author} {\bibfnamefont {J.}~\bibnamefont {Twamley}},\ }\bibfield  {title}
  {\emph {\bibinfo {title} {Magneto-mechanical trapping of micro-diamonds at
  low pressures},\ }}\href@noop {} {\bibfield  {journal} {\bibinfo  {journal}
  {Appl. Phys. Lett.}\ }\textbf {\bibinfo {volume} {114}},\ \bibinfo {pages}
  {053103} (\bibinfo {year} {2019})}\BibitemShut {NoStop}%
\bibitem [{\citenamefont {Hofer}\ and\ \citenamefont
  {Aspelmeyer}(2019)}]{hofer2019}%
  \BibitemOpen
  \bibfield  {author} {\bibinfo {author} {\bibfnamefont {J.}~\bibnamefont
  {Hofer}}\ and\ \bibinfo {author} {\bibfnamefont {M.}~\bibnamefont
  {Aspelmeyer}},\ }\bibfield  {title} {\emph {\bibinfo {title} {Analytic
  solutions to the {M}axwell--{L}ondon equations and levitation force for a
  superconducting sphere in a quadrupole field},\ }}\href@noop {} {\bibfield
  {journal} {\bibinfo  {journal} {Physica Scripta}\ }\textbf {\bibinfo {volume}
  {94}},\ \bibinfo {pages} {125508} (\bibinfo {year} {2019})}\BibitemShut
  {NoStop}%
\bibitem [{\citenamefont {Latorre}\ \emph {et~al.}(2020)\citenamefont
  {Latorre}, \citenamefont {Hofer}, \citenamefont {Rudolph},\ and\
  \citenamefont {Wieczorek}}]{latorre2020chip}%
  \BibitemOpen
  \bibfield  {author} {\bibinfo {author} {\bibfnamefont {M.~G.}\ \bibnamefont
  {Latorre}}, \bibinfo {author} {\bibfnamefont {J.}~\bibnamefont {Hofer}},
  \bibinfo {author} {\bibfnamefont {M.}~\bibnamefont {Rudolph}}, \ and\
  \bibinfo {author} {\bibfnamefont {W.}~\bibnamefont {Wieczorek}},\ }\bibfield
  {title} {\emph {\bibinfo {title} {Chip-based superconducting traps for
  levitation of micrometer-sized particles in the {M}eissner state},\
  }}\href@noop {} {\bibfield  {journal} {\bibinfo  {journal} {Superconductor
  Science and Technology}\ }\textbf {\bibinfo {volume} {33}},\ \bibinfo {pages}
  {105002} (\bibinfo {year} {2020})}\BibitemShut {NoStop}%
\bibitem [{\citenamefont {Druge}\ \emph {et~al.}(2014)\citenamefont {Druge},
  \citenamefont {Jean}, \citenamefont {Laurent}, \citenamefont {M{\'e}asson},\
  and\ \citenamefont {Favero}}]{druge2014damping}%
  \BibitemOpen
  \bibfield  {author} {\bibinfo {author} {\bibfnamefont {J.}~\bibnamefont
  {Druge}}, \bibinfo {author} {\bibfnamefont {C.}~\bibnamefont {Jean}},
  \bibinfo {author} {\bibfnamefont {O.}~\bibnamefont {Laurent}}, \bibinfo
  {author} {\bibfnamefont {M.-A.}\ \bibnamefont {M{\'e}asson}}, \ and\ \bibinfo
  {author} {\bibfnamefont {I.}~\bibnamefont {Favero}},\ }\bibfield  {title}
  {\emph {\bibinfo {title} {Damping and non-linearity of a levitating magnet in
  rotation above a superconductor},\ }}\href@noop {} {\bibfield  {journal}
  {\bibinfo  {journal} {New Journal of Physics}\ }\textbf {\bibinfo {volume}
  {16}},\ \bibinfo {pages} {075011} (\bibinfo {year} {2014})}\BibitemShut
  {NoStop}%
\bibitem [{\citenamefont {Timberlake}\ \emph {et~al.}(2019)\citenamefont
  {Timberlake}, \citenamefont {Gasbarri}, \citenamefont {Vinante},
  \citenamefont {Setter},\ and\ \citenamefont {Ulbricht}}]{timberlake2019}%
  \BibitemOpen
  \bibfield  {author} {\bibinfo {author} {\bibfnamefont {C.}~\bibnamefont
  {Timberlake}}, \bibinfo {author} {\bibfnamefont {G.}~\bibnamefont
  {Gasbarri}}, \bibinfo {author} {\bibfnamefont {A.}~\bibnamefont {Vinante}},
  \bibinfo {author} {\bibfnamefont {A.}~\bibnamefont {Setter}}, \ and\ \bibinfo
  {author} {\bibfnamefont {H.}~\bibnamefont {Ulbricht}},\ }\bibfield  {title}
  {\emph {\bibinfo {title} {Acceleration sensing with magnetically levitated
  oscillators above a superconductor},\ }}\href@noop {} {\bibfield  {journal}
  {\bibinfo  {journal} {Appl. Phys. Lett.}\ }\textbf {\bibinfo {volume}
  {115}},\ \bibinfo {pages} {224101} (\bibinfo {year} {2019})}\BibitemShut
  {NoStop}%
\bibitem [{\citenamefont {Wang}\ \emph {et~al.}(2019)\citenamefont {Wang},
  \citenamefont {Lourette}, \citenamefont {O'Kelley}, \citenamefont {Kayci},
  \citenamefont {Band}, \citenamefont {Kimball}, \citenamefont {Sushkov},\ and\
  \citenamefont {Budker}}]{wang2019}%
  \BibitemOpen
  \bibfield  {author} {\bibinfo {author} {\bibfnamefont {T.}~\bibnamefont
  {Wang}}, \bibinfo {author} {\bibfnamefont {S.}~\bibnamefont {Lourette}},
  \bibinfo {author} {\bibfnamefont {S.~R.}\ \bibnamefont {O'Kelley}}, \bibinfo
  {author} {\bibfnamefont {M.}~\bibnamefont {Kayci}}, \bibinfo {author}
  {\bibfnamefont {Y.}~\bibnamefont {Band}}, \bibinfo {author} {\bibfnamefont
  {D.~F.~J.}\ \bibnamefont {Kimball}}, \bibinfo {author} {\bibfnamefont
  {A.~O.}\ \bibnamefont {Sushkov}}, \ and\ \bibinfo {author} {\bibfnamefont
  {D.}~\bibnamefont {Budker}},\ }\bibfield  {title} {\emph {\bibinfo {title}
  {Dynamics of a ferromagnetic particle levitated over a superconductor},\
  }}\href {\doibase 10.1103/PhysRevApplied.11.044041} {\bibfield  {journal}
  {\bibinfo  {journal} {Phys. Rev. Applied}\ }\textbf {\bibinfo {volume}
  {11}},\ \bibinfo {pages} {044041} (\bibinfo {year} {2019})}\BibitemShut
  {NoStop}%
\bibitem [{\citenamefont {Vinante}\ \emph {et~al.}(2020)\citenamefont
  {Vinante}, \citenamefont {Falferi}, \citenamefont {Gasbarri}, \citenamefont
  {Setter}, \citenamefont {Timberlake},\ and\ \citenamefont
  {Ulbricht}}]{vinante2020}%
  \BibitemOpen
  \bibfield  {author} {\bibinfo {author} {\bibfnamefont {A.}~\bibnamefont
  {Vinante}}, \bibinfo {author} {\bibfnamefont {P.}~\bibnamefont {Falferi}},
  \bibinfo {author} {\bibfnamefont {G.}~\bibnamefont {Gasbarri}}, \bibinfo
  {author} {\bibfnamefont {A.}~\bibnamefont {Setter}}, \bibinfo {author}
  {\bibfnamefont {C.}~\bibnamefont {Timberlake}}, \ and\ \bibinfo {author}
  {\bibfnamefont {H.}~\bibnamefont {Ulbricht}},\ }\bibfield  {title} {\emph
  {\bibinfo {title} {Ultralow mechanical damping with {M}eissner-levitated
  ferromagnetic microparticles},\ }}\href {\doibase
  10.1103/PhysRevApplied.13.064027} {\bibfield  {journal} {\bibinfo  {journal}
  {Phys. Rev. Applied}\ }\textbf {\bibinfo {volume} {13}},\ \bibinfo {pages}
  {064027} (\bibinfo {year} {2020})}\BibitemShut {NoStop}%
\bibitem [{\citenamefont {Gieseler}\ \emph
  {et~al.}(2020{\natexlab{b}})\citenamefont {Gieseler}, \citenamefont
  {Kabcenell}, \citenamefont {Rosenfeld}, \citenamefont {Schaefer},
  \citenamefont {Safira}, \citenamefont {Schuetz}, \citenamefont
  {Gonzalez-Ballestero}, \citenamefont {Rusconi}, \citenamefont
  {Romero-Isart},\ and\ \citenamefont {Lukin}}]{gieseler2020b}%
  \BibitemOpen
  \bibfield  {author} {\bibinfo {author} {\bibfnamefont {J.}~\bibnamefont
  {Gieseler}}, \bibinfo {author} {\bibfnamefont {A.}~\bibnamefont {Kabcenell}},
  \bibinfo {author} {\bibfnamefont {E.}~\bibnamefont {Rosenfeld}}, \bibinfo
  {author} {\bibfnamefont {J.~D.}\ \bibnamefont {Schaefer}}, \bibinfo {author}
  {\bibfnamefont {A.}~\bibnamefont {Safira}}, \bibinfo {author} {\bibfnamefont
  {M.~J.~A.}\ \bibnamefont {Schuetz}}, \bibinfo {author} {\bibfnamefont
  {C.}~\bibnamefont {Gonzalez-Ballestero}}, \bibinfo {author} {\bibfnamefont
  {C.~C.}\ \bibnamefont {Rusconi}}, \bibinfo {author} {\bibfnamefont
  {O.}~\bibnamefont {Romero-Isart}}, \ and\ \bibinfo {author} {\bibfnamefont
  {M.~D.}\ \bibnamefont {Lukin}},\ }\bibfield  {title} {\emph {\bibinfo {title}
  {Single-spin magnetomechanics with levitated micromagnets},\ }}\href
  {\doibase 10.1103/PhysRevLett.124.163604} {\bibfield  {journal} {\bibinfo
  {journal} {Phys. Rev. Lett.}\ }\textbf {\bibinfo {volume} {124}},\ \bibinfo
  {pages} {163604} (\bibinfo {year} {2020}{\natexlab{b}})}\BibitemShut
  {NoStop}%
\bibitem [{\citenamefont {Prat-Camps}\ \emph {et~al.}(2017)\citenamefont
  {Prat-Camps}, \citenamefont {Teo}, \citenamefont {Rusconi}, \citenamefont
  {Wieczorek},\ and\ \citenamefont {Romero-Isart}}]{pratcamps2017}%
  \BibitemOpen
  \bibfield  {author} {\bibinfo {author} {\bibfnamefont {J.}~\bibnamefont
  {Prat-Camps}}, \bibinfo {author} {\bibfnamefont {C.}~\bibnamefont {Teo}},
  \bibinfo {author} {\bibfnamefont {C.~C.}\ \bibnamefont {Rusconi}}, \bibinfo
  {author} {\bibfnamefont {W.}~\bibnamefont {Wieczorek}}, \ and\ \bibinfo
  {author} {\bibfnamefont {O.}~\bibnamefont {Romero-Isart}},\ }\bibfield
  {title} {\emph {\bibinfo {title} {Ultrasensitive inertial and force sensors
  with diamagnetically levitated magnets},\ }}\href {\doibase
  10.1103/PhysRevApplied.8.034002} {\bibfield  {journal} {\bibinfo  {journal}
  {Phys. Rev. Applied}\ }\textbf {\bibinfo {volume} {8}},\ \bibinfo {pages}
  {034002} (\bibinfo {year} {2017})}\BibitemShut {NoStop}%
\bibitem [{\citenamefont {Kordyuk}(1998)}]{kordyuk1998}%
  \BibitemOpen
  \bibfield  {author} {\bibinfo {author} {\bibfnamefont {A.~A.}\ \bibnamefont
  {Kordyuk}},\ }\bibfield  {title} {\emph {\bibinfo {title} {Magnetic
  levitation for hard superconductors},\ }}\href@noop {} {\bibfield  {journal}
  {\bibinfo  {journal} {J. Appl. Phys.}\ }\textbf {\bibinfo {volume} {83}},\
  \bibinfo {pages} {610} (\bibinfo {year} {1998})}\BibitemShut {NoStop}%
\bibitem [{\citenamefont {Rusconi}\ and\ \citenamefont
  {Romero-Isart}(2016)}]{rusconi2016}%
  \BibitemOpen
  \bibfield  {author} {\bibinfo {author} {\bibfnamefont {C.~C.}\ \bibnamefont
  {Rusconi}}\ and\ \bibinfo {author} {\bibfnamefont {O.}~\bibnamefont
  {Romero-Isart}},\ }\bibfield  {title} {\emph {\bibinfo {title} {Magnetic
  rigid rotor in the quantum regime: Theoretical toolbox},\ }}\href {\doibase
  10.1103/PhysRevB.93.054427} {\bibfield  {journal} {\bibinfo  {journal} {Phys.
  Rev. B}\ }\textbf {\bibinfo {volume} {93}},\ \bibinfo {pages} {054427}
  (\bibinfo {year} {2016})}\BibitemShut {NoStop}%
\bibitem [{\citenamefont {Rusconi}\ \emph
  {et~al.}(2017{\natexlab{b}})\citenamefont {Rusconi}, \citenamefont
  {P{\"o}chhacker}, \citenamefont {Cirac},\ and\ \citenamefont
  {Romero-Isart}}]{rusconi2017b}%
  \BibitemOpen
  \bibfield  {author} {\bibinfo {author} {\bibfnamefont {C.~C.}\ \bibnamefont
  {Rusconi}}, \bibinfo {author} {\bibfnamefont {V.}~\bibnamefont
  {P{\"o}chhacker}}, \bibinfo {author} {\bibfnamefont {J.~I.}\ \bibnamefont
  {Cirac}}, \ and\ \bibinfo {author} {\bibfnamefont {O.}~\bibnamefont
  {Romero-Isart}},\ }\bibfield  {title} {\emph {\bibinfo {title} {Linear
  stability analysis of a levitated nanomagnet in a static magnetic field:
  Quantum spin stabilized magnetic levitation},\ }}\href@noop {} {\bibfield
  {journal} {\bibinfo  {journal} {Phys. Rev. B}\ }\textbf {\bibinfo {volume}
  {96}},\ \bibinfo {pages} {134419} (\bibinfo {year}
  {2017}{\natexlab{b}})}\BibitemShut {NoStop}%
\bibitem [{\citenamefont {Chang}\ \emph {et~al.}(2010)\citenamefont {Chang},
  \citenamefont {Regal}, \citenamefont {Papp}, \citenamefont {Wilson},
  \citenamefont {Ye}, \citenamefont {Painter}, \citenamefont {Kimble},\ and\
  \citenamefont {Zoller}}]{chang2010}%
  \BibitemOpen
  \bibfield  {author} {\bibinfo {author} {\bibfnamefont {D.}~\bibnamefont
  {Chang}}, \bibinfo {author} {\bibfnamefont {C.}~\bibnamefont {Regal}},
  \bibinfo {author} {\bibfnamefont {S.}~\bibnamefont {Papp}}, \bibinfo {author}
  {\bibfnamefont {D.}~\bibnamefont {Wilson}}, \bibinfo {author} {\bibfnamefont
  {J.}~\bibnamefont {Ye}}, \bibinfo {author} {\bibfnamefont {O.}~\bibnamefont
  {Painter}}, \bibinfo {author} {\bibfnamefont {H.}~\bibnamefont {Kimble}}, \
  and\ \bibinfo {author} {\bibfnamefont {P.}~\bibnamefont {Zoller}},\
  }\bibfield  {title} {\emph {\bibinfo {title} {{Cavity opto-mechanics using an
  optically levitated nanosphere}},\ }}\href@noop {} {\bibfield  {journal}
  {\bibinfo  {journal} {Proc. Natl. Acad. Sci. USA}\ }\textbf {\bibinfo
  {volume} {107}},\ \bibinfo {pages} {1005} (\bibinfo {year}
  {2010})}\BibitemShut {NoStop}%
\bibitem [{\citenamefont {Romero-Isart}\ \emph {et~al.}(2010)\citenamefont
  {Romero-Isart}, \citenamefont {Juan}, \citenamefont {Quidant},\ and\
  \citenamefont {Cirac}}]{romeroisart2010}%
  \BibitemOpen
  \bibfield  {author} {\bibinfo {author} {\bibfnamefont {O.}~\bibnamefont
  {Romero-Isart}}, \bibinfo {author} {\bibfnamefont {M.~L.}\ \bibnamefont
  {Juan}}, \bibinfo {author} {\bibfnamefont {R.}~\bibnamefont {Quidant}}, \
  and\ \bibinfo {author} {\bibfnamefont {J.~I.}\ \bibnamefont {Cirac}},\
  }\bibfield  {title} {\emph {\bibinfo {title} {Toward quantum superposition of
  living organisms},\ }}\href@noop {} {\bibfield  {journal} {\bibinfo
  {journal} {New J. Phys.}\ }\textbf {\bibinfo {volume} {12}},\ \bibinfo
  {pages} {033015} (\bibinfo {year} {2010})}\BibitemShut {NoStop}%
\bibitem [{\citenamefont {Barker}\ and\ \citenamefont
  {Shneider}(2010)}]{barker2010}%
  \BibitemOpen
  \bibfield  {author} {\bibinfo {author} {\bibfnamefont {P.}~\bibnamefont
  {Barker}}\ and\ \bibinfo {author} {\bibfnamefont {M.}~\bibnamefont
  {Shneider}},\ }\bibfield  {title} {\emph {\bibinfo {title} {Cavity cooling of
  an optically trapped nanoparticle},\ }}\href@noop {} {\bibfield  {journal}
  {\bibinfo  {journal} {Phys. Rev. A}\ }\textbf {\bibinfo {volume} {81}},\
  \bibinfo {pages} {023826} (\bibinfo {year} {2010})}\BibitemShut {NoStop}%
\bibitem [{\citenamefont {Kiesel}\ \emph {et~al.}(2013)\citenamefont {Kiesel},
  \citenamefont {Blaser}, \citenamefont {Deli{\'c}}, \citenamefont {Grass},
  \citenamefont {Kaltenbaek},\ and\ \citenamefont {Aspelmeyer}}]{kiesel2013}%
  \BibitemOpen
  \bibfield  {author} {\bibinfo {author} {\bibfnamefont {N.}~\bibnamefont
  {Kiesel}}, \bibinfo {author} {\bibfnamefont {F.}~\bibnamefont {Blaser}},
  \bibinfo {author} {\bibfnamefont {U.}~\bibnamefont {Deli{\'c}}}, \bibinfo
  {author} {\bibfnamefont {D.}~\bibnamefont {Grass}}, \bibinfo {author}
  {\bibfnamefont {R.}~\bibnamefont {Kaltenbaek}}, \ and\ \bibinfo {author}
  {\bibfnamefont {M.}~\bibnamefont {Aspelmeyer}},\ }\bibfield  {title} {\emph
  {\bibinfo {title} {Cavity cooling of an optically levitated submicron
  particle},\ }}\href@noop {} {\bibfield  {journal} {\bibinfo  {journal} {Proc.
  Natl. Acad. Sci. USA}\ }\textbf {\bibinfo {volume} {110}},\ \bibinfo {pages}
  {14180} (\bibinfo {year} {2013})}\BibitemShut {NoStop}%
\bibitem [{\citenamefont {Asenbaum}\ \emph {et~al.}(2013)\citenamefont
  {Asenbaum}, \citenamefont {Kuhn}, \citenamefont {Nimmrichter}, \citenamefont
  {Sezer},\ and\ \citenamefont {Arndt}}]{asenbaum2013}%
  \BibitemOpen
  \bibfield  {author} {\bibinfo {author} {\bibfnamefont {P.}~\bibnamefont
  {Asenbaum}}, \bibinfo {author} {\bibfnamefont {S.}~\bibnamefont {Kuhn}},
  \bibinfo {author} {\bibfnamefont {S.}~\bibnamefont {Nimmrichter}}, \bibinfo
  {author} {\bibfnamefont {U.}~\bibnamefont {Sezer}}, \ and\ \bibinfo {author}
  {\bibfnamefont {M.}~\bibnamefont {Arndt}},\ }\bibfield  {title} {\emph
  {\bibinfo {title} {Cavity cooling of free silicon nanoparticles in high
  vacuum},\ }}\href@noop {} {\bibfield  {journal} {\bibinfo  {journal} {Nat.
  Commun.}\ }\textbf {\bibinfo {volume} {4}},\ \bibinfo {pages} {2743}
  (\bibinfo {year} {2013})}\BibitemShut {NoStop}%
\bibitem [{\citenamefont {Fonseca}\ \emph {et~al.}(2016)\citenamefont
  {Fonseca}, \citenamefont {Aranas}, \citenamefont {Millen}, \citenamefont
  {Monteiro},\ and\ \citenamefont {Barker}}]{fonseca2016}%
  \BibitemOpen
  \bibfield  {author} {\bibinfo {author} {\bibfnamefont {P.~Z.~G.}\
  \bibnamefont {Fonseca}}, \bibinfo {author} {\bibfnamefont {E.~B.}\
  \bibnamefont {Aranas}}, \bibinfo {author} {\bibfnamefont {J.}~\bibnamefont
  {Millen}}, \bibinfo {author} {\bibfnamefont {T.~S.}\ \bibnamefont
  {Monteiro}}, \ and\ \bibinfo {author} {\bibfnamefont {P.~F.}\ \bibnamefont
  {Barker}},\ }\bibfield  {title} {\emph {\bibinfo {title} {Nonlinear dynamics
  and strong cavity cooling of levitated nanoparticles},\ }}\href@noop {}
  {\bibfield  {journal} {\bibinfo  {journal} {Phys. Rev. Lett.}\ }\textbf
  {\bibinfo {volume} {117}},\ \bibinfo {pages} {173602} (\bibinfo {year}
  {2016})}\BibitemShut {NoStop}%
\bibitem [{\citenamefont {Salzburger}\ and\ \citenamefont
  {Ritsch}(2009)}]{ritschnjp2009}%
  \BibitemOpen
  \bibfield  {author} {\bibinfo {author} {\bibfnamefont {T.}~\bibnamefont
  {Salzburger}}\ and\ \bibinfo {author} {\bibfnamefont {H.}~\bibnamefont
  {Ritsch}},\ }\bibfield  {title} {\emph {\bibinfo {title} {{Collective
  transverse cavity cooling of a dense molecular beam}},\ }}\href
  {http://stacks.iop.org/1367-2630/11/i=5/a=055025} {\bibfield  {journal}
  {\bibinfo  {journal} {New J. Phys.}\ }\textbf {\bibinfo {volume} {11}},\
  \bibinfo {pages} {055025} (\bibinfo {year} {2009})}\BibitemShut {NoStop}%
\bibitem [{\citenamefont {Gonzalez-Ballestero}\ \emph
  {et~al.}(2019)\citenamefont {Gonzalez-Ballestero}, \citenamefont {Maurer},
  \citenamefont {Windey}, \citenamefont {Novotny}, \citenamefont {Reimann},\
  and\ \citenamefont {Romero-Isart}}]{gonzalezballestero2019}%
  \BibitemOpen
  \bibfield  {author} {\bibinfo {author} {\bibfnamefont {C.}~\bibnamefont
  {Gonzalez-Ballestero}}, \bibinfo {author} {\bibfnamefont {P.}~\bibnamefont
  {Maurer}}, \bibinfo {author} {\bibfnamefont {D.}~\bibnamefont {Windey}},
  \bibinfo {author} {\bibfnamefont {L.}~\bibnamefont {Novotny}}, \bibinfo
  {author} {\bibfnamefont {R.}~\bibnamefont {Reimann}}, \ and\ \bibinfo
  {author} {\bibfnamefont {O.}~\bibnamefont {Romero-Isart}},\ }\bibfield
  {title} {\emph {\bibinfo {title} {Theory for cavity cooling of levitated
  nanoparticles via coherent scattering: Master equation approach},\ }}\href
  {\doibase 10.1103/PhysRevA.100.013805} {\bibfield  {journal} {\bibinfo
  {journal} {Phys. Rev. A}\ }\textbf {\bibinfo {volume} {100}},\ \bibinfo
  {pages} {013805} (\bibinfo {year} {2019})}\BibitemShut {NoStop}%
\bibitem [{\citenamefont {Windey}\ \emph {et~al.}(2019)\citenamefont {Windey},
  \citenamefont {Gonzalez-Ballestero}, \citenamefont {Maurer}, \citenamefont
  {Novotny}, \citenamefont {Romero-Isart},\ and\ \citenamefont
  {Reimann}}]{windey2019}%
  \BibitemOpen
  \bibfield  {author} {\bibinfo {author} {\bibfnamefont {D.}~\bibnamefont
  {Windey}}, \bibinfo {author} {\bibfnamefont {C.}~\bibnamefont
  {Gonzalez-Ballestero}}, \bibinfo {author} {\bibfnamefont {P.}~\bibnamefont
  {Maurer}}, \bibinfo {author} {\bibfnamefont {L.}~\bibnamefont {Novotny}},
  \bibinfo {author} {\bibfnamefont {O.}~\bibnamefont {Romero-Isart}}, \ and\
  \bibinfo {author} {\bibfnamefont {R.}~\bibnamefont {Reimann}},\ }\bibfield
  {title} {\emph {\bibinfo {title} {Cavity-based {3D} cooling of a levitated
  nanoparticle via coherent scattering},\ }}\href {\doibase
  10.1103/PhysRevLett.122.123601} {\bibfield  {journal} {\bibinfo  {journal}
  {Phys. Rev. Lett.}\ }\textbf {\bibinfo {volume} {122}},\ \bibinfo {pages}
  {123601} (\bibinfo {year} {2019})}\BibitemShut {NoStop}%
\bibitem [{\citenamefont {Deli\v{c}}\ \emph {et~al.}(2019)\citenamefont
  {Deli\v{c}}, \citenamefont {Reisenbauer}, \citenamefont {Grass},
  \citenamefont {Kiesel}, \citenamefont {Vuleti\v{c}},\ and\ \citenamefont
  {Aspelmeyer}}]{delic2019}%
  \BibitemOpen
  \bibfield  {author} {\bibinfo {author} {\bibfnamefont {U.}~\bibnamefont
  {Deli\v{c}}}, \bibinfo {author} {\bibfnamefont {M.}~\bibnamefont
  {Reisenbauer}}, \bibinfo {author} {\bibfnamefont {D.}~\bibnamefont {Grass}},
  \bibinfo {author} {\bibfnamefont {N.}~\bibnamefont {Kiesel}}, \bibinfo
  {author} {\bibfnamefont {V.}~\bibnamefont {Vuleti\v{c}}}, \ and\ \bibinfo
  {author} {\bibfnamefont {M.}~\bibnamefont {Aspelmeyer}},\ }\bibfield  {title}
  {\emph {\bibinfo {title} {Cavity cooling of a levitated nanosphere by
  coherent scattering},\ }}\href {\doibase 10.1103/PhysRevLett.122.123602}
  {\bibfield  {journal} {\bibinfo  {journal} {Phys. Rev. Lett.}\ }\textbf
  {\bibinfo {volume} {122}},\ \bibinfo {pages} {123602} (\bibinfo {year}
  {2019})}\BibitemShut {NoStop}%
\bibitem [{\citenamefont {Schlosshauer}(2019)}]{schlosshauer2019}%
  \BibitemOpen
  \bibfield  {author} {\bibinfo {author} {\bibfnamefont {M.}~\bibnamefont
  {Schlosshauer}},\ }\bibfield  {title} {\emph {\bibinfo {title} {Quantum
  decoherence},\ }}\href@noop {} {\bibfield  {journal} {\bibinfo  {journal}
  {Phys. Rep.}\ }\textbf {\bibinfo {volume} {831}},\ \bibinfo {pages} {1}
  (\bibinfo {year} {2019})}\BibitemShut {NoStop}%
\bibitem [{\citenamefont {Stickler}\ \emph
  {et~al.}(2016{\natexlab{b}})\citenamefont {Stickler}, \citenamefont
  {Papendell},\ and\ \citenamefont {Hornberger}}]{stickler2016b}%
  \BibitemOpen
  \bibfield  {author} {\bibinfo {author} {\bibfnamefont {B.~A.}\ \bibnamefont
  {Stickler}}, \bibinfo {author} {\bibfnamefont {B.}~\bibnamefont {Papendell}},
  \ and\ \bibinfo {author} {\bibfnamefont {K.}~\bibnamefont {Hornberger}},\
  }\bibfield  {title} {\emph {\bibinfo {title} {Spatio-orientational
  decoherence of nanoparticles},\ }}\href@noop {} {\bibfield  {journal}
  {\bibinfo  {journal} {Phys. Rev. A}\ }\textbf {\bibinfo {volume} {94}},\
  \bibinfo {pages} {033828} (\bibinfo {year} {2016}{\natexlab{b}})}\BibitemShut
  {NoStop}%
\bibitem [{\citenamefont {Zhong}\ and\ \citenamefont
  {Robicheaux}(2016)}]{zhong2016}%
  \BibitemOpen
  \bibfield  {author} {\bibinfo {author} {\bibfnamefont {C.}~\bibnamefont
  {Zhong}}\ and\ \bibinfo {author} {\bibfnamefont {F.}~\bibnamefont
  {Robicheaux}},\ }\bibfield  {title} {\emph {\bibinfo {title} {Decoherence of
  rotational degrees of freedom},\ }}\href@noop {} {\bibfield  {journal}
  {\bibinfo  {journal} {Phys. Rev. A}\ }\textbf {\bibinfo {volume} {94}},\
  \bibinfo {pages} {052109} (\bibinfo {year} {2016})}\BibitemShut {NoStop}%
\bibitem [{\citenamefont {Papendell}\ \emph {et~al.}(2017)\citenamefont
  {Papendell}, \citenamefont {Stickler},\ and\ \citenamefont
  {Hornberger}}]{papendell2017}%
  \BibitemOpen
  \bibfield  {author} {\bibinfo {author} {\bibfnamefont {B.}~\bibnamefont
  {Papendell}}, \bibinfo {author} {\bibfnamefont {B.~A.}\ \bibnamefont
  {Stickler}}, \ and\ \bibinfo {author} {\bibfnamefont {K.}~\bibnamefont
  {Hornberger}},\ }\bibfield  {title} {\emph {\bibinfo {title} {Quantum angular
  momentum diffusion of rigid bodies},\ }}\href
  {http://stacks.iop.org/1367-2630/19/i=12/a=122001} {\bibfield  {journal}
  {\bibinfo  {journal} {New J. Phys.}\ }\textbf {\bibinfo {volume} {19}},\
  \bibinfo {pages} {122001} (\bibinfo {year} {2017})}\BibitemShut {NoStop}%
\bibitem [{\citenamefont {Pedernales}\ \emph {et~al.}(2020)\citenamefont
  {Pedernales}, \citenamefont {Cosco},\ and\ \citenamefont
  {Plenio}}]{pedernales2020b}%
  \BibitemOpen
  \bibfield  {author} {\bibinfo {author} {\bibfnamefont {J.~S.}\ \bibnamefont
  {Pedernales}}, \bibinfo {author} {\bibfnamefont {F.}~\bibnamefont {Cosco}}, \
  and\ \bibinfo {author} {\bibfnamefont {M.~B.}\ \bibnamefont {Plenio}},\
  }\bibfield  {title} {\emph {\bibinfo {title} {Decoherence-free rotational
  degrees of freedom for quantum applications},\ }}\href {\doibase
  10.1103/PhysRevLett.125.090501} {\bibfield  {journal} {\bibinfo  {journal}
  {Phys. Rev. Lett.}\ }\textbf {\bibinfo {volume} {125}},\ \bibinfo {pages}
  {090501} (\bibinfo {year} {2020})}\BibitemShut {NoStop}%
\bibitem [{\citenamefont {Seberson}\ and\ \citenamefont
  {Robicheaux}(2020{\natexlab{a}})}]{seberson2020distribution}%
  \BibitemOpen
  \bibfield  {author} {\bibinfo {author} {\bibfnamefont {T.}~\bibnamefont
  {Seberson}}\ and\ \bibinfo {author} {\bibfnamefont {F.}~\bibnamefont
  {Robicheaux}},\ }\bibfield  {title} {\emph {\bibinfo {title} {Distribution of
  laser shot-noise energy delivered to a levitated nanoparticle},\ }}\href
  {\doibase 10.1103/PhysRevA.102.033505} {\bibfield  {journal} {\bibinfo
  {journal} {Phys. Rev. A}\ }\textbf {\bibinfo {volume} {102}},\ \bibinfo
  {pages} {033505} (\bibinfo {year} {2020}{\natexlab{a}})}\BibitemShut
  {NoStop}%
\bibitem [{\citenamefont {Stickler}\ \emph
  {et~al.}(2018{\natexlab{b}})\citenamefont {Stickler}, \citenamefont
  {Schrinski},\ and\ \citenamefont {Hornberger}}]{stickler2018a}%
  \BibitemOpen
  \bibfield  {author} {\bibinfo {author} {\bibfnamefont {B.~A.}\ \bibnamefont
  {Stickler}}, \bibinfo {author} {\bibfnamefont {B.}~\bibnamefont {Schrinski}},
  \ and\ \bibinfo {author} {\bibfnamefont {K.}~\bibnamefont {Hornberger}},\
  }\bibfield  {title} {\emph {\bibinfo {title} {Rotational friction and
  diffusion of quantum rotors},\ }}\href {\doibase
  10.1103/PhysRevLett.121.040401} {\bibfield  {journal} {\bibinfo  {journal}
  {Phys. Rev. Lett.}\ }\textbf {\bibinfo {volume} {121}},\ \bibinfo {pages}
  {040401} (\bibinfo {year} {2018}{\natexlab{b}})}\BibitemShut {NoStop}%
\bibitem [{\citenamefont {Yin}\ \emph {et~al.}(2013)\citenamefont {Yin},
  \citenamefont {Li}, \citenamefont {Zhang},\ and\ \citenamefont
  {Duan}}]{yin2013}%
  \BibitemOpen
  \bibfield  {author} {\bibinfo {author} {\bibfnamefont {Z.-q.}\ \bibnamefont
  {Yin}}, \bibinfo {author} {\bibfnamefont {T.}~\bibnamefont {Li}}, \bibinfo
  {author} {\bibfnamefont {X.}~\bibnamefont {Zhang}}, \ and\ \bibinfo {author}
  {\bibfnamefont {L.}~\bibnamefont {Duan}},\ }\bibfield  {title} {\emph
  {\bibinfo {title} {Large quantum superpositions of a levitated nanodiamond
  through spin-optomechanical coupling},\ }}\href@noop {} {\bibfield  {journal}
  {\bibinfo  {journal} {Phys. Rev. A}\ }\textbf {\bibinfo {volume} {88}},\
  \bibinfo {pages} {033614} (\bibinfo {year} {2013})}\BibitemShut {NoStop}%
\bibitem [{\citenamefont {Pflanzer}\ \emph {et~al.}(2013)\citenamefont
  {Pflanzer}, \citenamefont {Romero-Isart},\ and\ \citenamefont
  {Cirac}}]{pflanzer2013}%
  \BibitemOpen
  \bibfield  {author} {\bibinfo {author} {\bibfnamefont {A.~C.}\ \bibnamefont
  {Pflanzer}}, \bibinfo {author} {\bibfnamefont {O.}~\bibnamefont
  {Romero-Isart}}, \ and\ \bibinfo {author} {\bibfnamefont {J.~I.}\
  \bibnamefont {Cirac}},\ }\bibfield  {title} {\emph {\bibinfo {title}
  {Optomechanics assisted by a qubit: from dissipative state preparation to
  many-partite systems},\ }}\href@noop {} {\bibfield  {journal} {\bibinfo
  {journal} {Phys. Rev. A}\ }\textbf {\bibinfo {volume} {88}},\ \bibinfo
  {pages} {033804} (\bibinfo {year} {2013})}\BibitemShut {NoStop}%
\bibitem [{\citenamefont {Delord}\ \emph
  {et~al.}(2017{\natexlab{c}})\citenamefont {Delord}, \citenamefont {Nicolas},
  \citenamefont {Bodini},\ and\ \citenamefont {H{\'e}tet}}]{delord2017c}%
  \BibitemOpen
  \bibfield  {author} {\bibinfo {author} {\bibfnamefont {T.}~\bibnamefont
  {Delord}}, \bibinfo {author} {\bibfnamefont {L.}~\bibnamefont {Nicolas}},
  \bibinfo {author} {\bibfnamefont {M.}~\bibnamefont {Bodini}}, \ and\ \bibinfo
  {author} {\bibfnamefont {G.}~\bibnamefont {H{\'e}tet}},\ }\bibfield  {title}
  {\emph {\bibinfo {title} {Diamonds levitating in a {Paul} trap under vacuum:
  Measurements of laser-induced heating via {NV} center thermometry},\
  }}\href@noop {} {\bibfield  {journal} {\bibinfo  {journal} {Appl. Phys.
  Lett.}\ }\textbf {\bibinfo {volume} {111}},\ \bibinfo {pages} {013101}
  (\bibinfo {year} {2017}{\natexlab{c}})}\BibitemShut {NoStop}%
\bibitem [{\citenamefont {Delord}\ \emph {et~al.}(2018)\citenamefont {Delord},
  \citenamefont {Huillery}, \citenamefont {Schwab}, \citenamefont {Nicolas},
  \citenamefont {Lecordier},\ and\ \citenamefont {H\'etet}}]{delord2018}%
  \BibitemOpen
  \bibfield  {author} {\bibinfo {author} {\bibfnamefont {T.}~\bibnamefont
  {Delord}}, \bibinfo {author} {\bibfnamefont {P.}~\bibnamefont {Huillery}},
  \bibinfo {author} {\bibfnamefont {L.}~\bibnamefont {Schwab}}, \bibinfo
  {author} {\bibfnamefont {L.}~\bibnamefont {Nicolas}}, \bibinfo {author}
  {\bibfnamefont {L.}~\bibnamefont {Lecordier}}, \ and\ \bibinfo {author}
  {\bibfnamefont {G.}~\bibnamefont {H\'etet}},\ }\bibfield  {title} {\emph
  {\bibinfo {title} {Ramsey interferences and spin echoes from electron spins
  inside a levitating macroscopic particle},\ }}\href {\doibase
  10.1103/PhysRevLett.121.053602} {\bibfield  {journal} {\bibinfo  {journal}
  {Phys. Rev. Lett.}\ }\textbf {\bibinfo {volume} {121}},\ \bibinfo {pages}
  {053602} (\bibinfo {year} {2018})}\BibitemShut {NoStop}%
\bibitem [{\citenamefont {Tebbenjohanns}\ \emph {et~al.}(2019)\citenamefont
  {Tebbenjohanns}, \citenamefont {Frimmer}, \citenamefont {Militaru},
  \citenamefont {Jain},\ and\ \citenamefont {Novotny}}]{tebbenjohanns2019}%
  \BibitemOpen
  \bibfield  {author} {\bibinfo {author} {\bibfnamefont {F.}~\bibnamefont
  {Tebbenjohanns}}, \bibinfo {author} {\bibfnamefont {M.}~\bibnamefont
  {Frimmer}}, \bibinfo {author} {\bibfnamefont {A.}~\bibnamefont {Militaru}},
  \bibinfo {author} {\bibfnamefont {V.}~\bibnamefont {Jain}}, \ and\ \bibinfo
  {author} {\bibfnamefont {L.}~\bibnamefont {Novotny}},\ }\bibfield  {title}
  {\emph {\bibinfo {title} {Cold damping of an optically levitated nanoparticle
  to microkelvin temperatures},\ }}\href {\doibase
  10.1103/PhysRevLett.122.223601} {\bibfield  {journal} {\bibinfo  {journal}
  {Phys. Rev. Lett.}\ }\textbf {\bibinfo {volume} {122}},\ \bibinfo {pages}
  {223601} (\bibinfo {year} {2019})}\BibitemShut {NoStop}%
\bibitem [{\citenamefont {Dania}\ \emph {et~al.}(2021)\citenamefont {Dania},
  \citenamefont {Bykov}, \citenamefont {Knoll}, \citenamefont {Mestres},\ and\
  \citenamefont {Northup}}]{dania2021}%
  \BibitemOpen
  \bibfield  {author} {\bibinfo {author} {\bibfnamefont {L.}~\bibnamefont
  {Dania}}, \bibinfo {author} {\bibfnamefont {D.~S.}\ \bibnamefont {Bykov}},
  \bibinfo {author} {\bibfnamefont {M.}~\bibnamefont {Knoll}}, \bibinfo
  {author} {\bibfnamefont {P.}~\bibnamefont {Mestres}}, \ and\ \bibinfo
  {author} {\bibfnamefont {T.~E.}\ \bibnamefont {Northup}},\ }\bibfield
  {title} {\emph {\bibinfo {title} {Optical and electrical feedback cooling of
  a silica nanoparticle levitated in a {Paul} trap},\ }}\href {\doibase
  10.1103/PhysRevResearch.3.013018} {\bibfield  {journal} {\bibinfo  {journal}
  {Phys. Rev. Research}\ }\textbf {\bibinfo {volume} {3}},\ \bibinfo {pages}
  {013018} (\bibinfo {year} {2021})}\BibitemShut {NoStop}%
\bibitem [{\citenamefont {Rider}\ \emph {et~al.}(2016)\citenamefont {Rider},
  \citenamefont {Moore}, \citenamefont {Blakemore}, \citenamefont {Louis},
  \citenamefont {Lu},\ and\ \citenamefont {Gratta}}]{rider2016}%
  \BibitemOpen
  \bibfield  {author} {\bibinfo {author} {\bibfnamefont {A.~D.}\ \bibnamefont
  {Rider}}, \bibinfo {author} {\bibfnamefont {D.~C.}\ \bibnamefont {Moore}},
  \bibinfo {author} {\bibfnamefont {C.~P.}\ \bibnamefont {Blakemore}}, \bibinfo
  {author} {\bibfnamefont {M.}~\bibnamefont {Louis}}, \bibinfo {author}
  {\bibfnamefont {M.}~\bibnamefont {Lu}}, \ and\ \bibinfo {author}
  {\bibfnamefont {G.}~\bibnamefont {Gratta}},\ }\bibfield  {title} {\emph
  {\bibinfo {title} {Search for screened interactions associated with dark
  energy below the 100 $\mu$ m length scale},\ }}\href@noop {} {\bibfield
  {journal} {\bibinfo  {journal} {Phys. Rev. Lett.}\ }\textbf {\bibinfo
  {volume} {117}},\ \bibinfo {pages} {101101} (\bibinfo {year}
  {2016})}\BibitemShut {NoStop}%
\bibitem [{\citenamefont {Jackson}(1999)}]{jackson1999}%
  \BibitemOpen
  \bibfield  {author} {\bibinfo {author} {\bibfnamefont {J.~D.}\ \bibnamefont
  {Jackson}},\ }\href@noop {} {\emph {\bibinfo {title} {{Classical
  Electrodynamics}}}}\ (\bibinfo  {publisher} {Wiley},\ \bibinfo {address} {New
  York},\ \bibinfo {year} {1999})\BibitemShut {NoStop}%
\bibitem [{\citenamefont {Jackson~Kimball}\ \emph {et~al.}(2016)\citenamefont
  {Jackson~Kimball}, \citenamefont {Sushkov},\ and\ \citenamefont
  {Budker}}]{kimball2016}%
  \BibitemOpen
  \bibfield  {author} {\bibinfo {author} {\bibfnamefont {D.~F.}\ \bibnamefont
  {Jackson~Kimball}}, \bibinfo {author} {\bibfnamefont {A.~O.}\ \bibnamefont
  {Sushkov}}, \ and\ \bibinfo {author} {\bibfnamefont {D.}~\bibnamefont
  {Budker}},\ }\bibfield  {title} {\emph {\bibinfo {title} {Precessing
  ferromagnetic needle magnetometer},\ }}\href {\doibase
  10.1103/PhysRevLett.116.190801} {\bibfield  {journal} {\bibinfo  {journal}
  {Phys. Rev. Lett.}\ }\textbf {\bibinfo {volume} {116}},\ \bibinfo {pages}
  {190801} (\bibinfo {year} {2016})}\BibitemShut {NoStop}%
\bibitem [{\citenamefont {Berry}(1996)}]{berry1996}%
  \BibitemOpen
  \bibfield  {author} {\bibinfo {author} {\bibfnamefont {M.~V.}\ \bibnamefont
  {Berry}},\ }\bibfield  {title} {\emph {\bibinfo {title} {The {Levitron}: an
  adiabatic trap for spins},\ }}\href@noop {} {\bibfield  {journal} {\bibinfo
  {journal} {Proceedings of the Royal Society of London. Series A:
  Mathematical, Physical and Engineering Sciences}\ }\textbf {\bibinfo {volume}
  {452}},\ \bibinfo {pages} {1207} (\bibinfo {year} {1996})}\BibitemShut
  {NoStop}%
\bibitem [{\citenamefont {Simon}\ \emph {et~al.}(1997)\citenamefont {Simon},
  \citenamefont {Heflinger},\ and\ \citenamefont {Ridgway}}]{simon1997spin}%
  \BibitemOpen
  \bibfield  {author} {\bibinfo {author} {\bibfnamefont {M.~D.}\ \bibnamefont
  {Simon}}, \bibinfo {author} {\bibfnamefont {L.~O.}\ \bibnamefont
  {Heflinger}}, \ and\ \bibinfo {author} {\bibfnamefont {S.}~\bibnamefont
  {Ridgway}},\ }\bibfield  {title} {\emph {\bibinfo {title} {Spin stabilized
  magnetic levitation},\ }}\href@noop {} {\bibfield  {journal} {\bibinfo
  {journal} {Am. J. Phys.}\ }\textbf {\bibinfo {volume} {65}},\ \bibinfo
  {pages} {286} (\bibinfo {year} {1997})}\BibitemShut {NoStop}%
\bibitem [{\citenamefont {Seberson}\ and\ \citenamefont
  {Robicheaux}(2020{\natexlab{b}})}]{seberson2020}%
  \BibitemOpen
  \bibfield  {author} {\bibinfo {author} {\bibfnamefont {T.}~\bibnamefont
  {Seberson}}\ and\ \bibinfo {author} {\bibfnamefont {F.}~\bibnamefont
  {Robicheaux}},\ }\bibfield  {title} {\emph {\bibinfo {title} {Stability and
  dynamics of optically levitated dielectric disks in a gaussian standing wave
  beyond the harmonic approximation},\ }}\href {\doibase
  10.1103/PhysRevResearch.2.033437} {\bibfield  {journal} {\bibinfo  {journal}
  {Phys. Rev. Research}\ }\textbf {\bibinfo {volume} {2}},\ \bibinfo {pages}
  {033437} (\bibinfo {year} {2020}{\natexlab{b}})}\BibitemShut {NoStop}%
\bibitem [{\citenamefont {Bowen}\ and\ \citenamefont
  {Milburn}(2015)}]{bowen2015}%
  \BibitemOpen
  \bibfield  {author} {\bibinfo {author} {\bibfnamefont {W.~P.}\ \bibnamefont
  {Bowen}}\ and\ \bibinfo {author} {\bibfnamefont {G.~J.}\ \bibnamefont
  {Milburn}},\ }\href@noop {} {\emph {\bibinfo {title} {Quantum
  optomechanics}}}\ (\bibinfo  {publisher} {CRC press},\ \bibinfo {year}
  {2015})\BibitemShut {NoStop}%
\bibitem [{\citenamefont {Xu}\ and\ \citenamefont {Li}(2017)}]{xu2017}%
  \BibitemOpen
  \bibfield  {author} {\bibinfo {author} {\bibfnamefont {Z.}~\bibnamefont
  {Xu}}\ and\ \bibinfo {author} {\bibfnamefont {T.}~\bibnamefont {Li}},\
  }\bibfield  {title} {\emph {\bibinfo {title} {Detecting casimir torque with
  an optically levitated nanorod},\ }}\href@noop {} {\bibfield  {journal}
  {\bibinfo  {journal} {Phys. Rev. A}\ }\textbf {\bibinfo {volume} {96}},\
  \bibinfo {pages} {033843} (\bibinfo {year} {2017})}\BibitemShut {NoStop}%
\bibitem [{\citenamefont {Zhao}\ \emph {et~al.}(2012)\citenamefont {Zhao},
  \citenamefont {Manjavacas}, \citenamefont {Garc\'{\i}a~de Abajo},\ and\
  \citenamefont {Pendry}}]{zhao2012}%
  \BibitemOpen
  \bibfield  {author} {\bibinfo {author} {\bibfnamefont {R.}~\bibnamefont
  {Zhao}}, \bibinfo {author} {\bibfnamefont {A.}~\bibnamefont {Manjavacas}},
  \bibinfo {author} {\bibfnamefont {F.~J.}\ \bibnamefont {Garc\'{\i}a~de
  Abajo}}, \ and\ \bibinfo {author} {\bibfnamefont {J.~B.}\ \bibnamefont
  {Pendry}},\ }\bibfield  {title} {\emph {\bibinfo {title} {Rotational quantum
  friction},\ }}\href {\doibase 10.1103/PhysRevLett.109.123604} {\bibfield
  {journal} {\bibinfo  {journal} {Phys. Rev. Lett.}\ }\textbf {\bibinfo
  {volume} {109}},\ \bibinfo {pages} {123604} (\bibinfo {year}
  {2012})}\BibitemShut {NoStop}%
\bibitem [{\citenamefont {Moore}\ and\ \citenamefont
  {Geraci}(2020)}]{moore2020}%
  \BibitemOpen
  \bibfield  {author} {\bibinfo {author} {\bibfnamefont {D.~C.}\ \bibnamefont
  {Moore}}\ and\ \bibinfo {author} {\bibfnamefont {A.}~\bibnamefont {Geraci}},\
  }\bibfield  {title} {\emph {\bibinfo {title} {Searching for new physics using
  optically levitated sensors},\ }}\href@noop {} {\bibfield  {journal}
  {\bibinfo  {journal} {Quant. Sci. Techn.}\ } (\bibinfo {year}
  {2020})}\BibitemShut {NoStop}%
\bibitem [{\citenamefont {Band}\ \emph {et~al.}(2018)\citenamefont {Band},
  \citenamefont {Avishai},\ and\ \citenamefont {Shnirman}}]{band2018}%
  \BibitemOpen
  \bibfield  {author} {\bibinfo {author} {\bibfnamefont {Y.~B.}\ \bibnamefont
  {Band}}, \bibinfo {author} {\bibfnamefont {Y.}~\bibnamefont {Avishai}}, \
  and\ \bibinfo {author} {\bibfnamefont {A.}~\bibnamefont {Shnirman}},\
  }\bibfield  {title} {\emph {\bibinfo {title} {Dynamics of a magnetic needle
  magnetometer: Sensitivity to {Landau-Lifshitz-Gilbert} damping},\ }}\href
  {\doibase 10.1103/PhysRevLett.121.160801} {\bibfield  {journal} {\bibinfo
  {journal} {Phys. Rev. Lett.}\ }\textbf {\bibinfo {volume} {121}},\ \bibinfo
  {pages} {160801} (\bibinfo {year} {2018})}\BibitemShut {NoStop}%
\bibitem [{\citenamefont {Bassi}\ \emph {et~al.}(2013)\citenamefont {Bassi},
  \citenamefont {Lochan}, \citenamefont {Satin}, \citenamefont {Singh},\ and\
  \citenamefont {Ulbricht}}]{bassi2013}%
  \BibitemOpen
  \bibfield  {author} {\bibinfo {author} {\bibfnamefont {A.}~\bibnamefont
  {Bassi}}, \bibinfo {author} {\bibfnamefont {K.}~\bibnamefont {Lochan}},
  \bibinfo {author} {\bibfnamefont {S.}~\bibnamefont {Satin}}, \bibinfo
  {author} {\bibfnamefont {T.}~\bibnamefont {Singh}}, \ and\ \bibinfo {author}
  {\bibfnamefont {H.}~\bibnamefont {Ulbricht}},\ }\bibfield  {title} {\emph
  {\bibinfo {title} {{Models of wave-function collapse, underlying theories,
  and experimental tests}},\ }}\href {\doibase 10.1103/RevModPhys.85.471}
  {\bibfield  {journal} {\bibinfo  {journal} {Rev. Mod. Phys.}\ }\textbf
  {\bibinfo {volume} {85}},\ \bibinfo {pages} {471} (\bibinfo {year}
  {2013})}\BibitemShut {NoStop}%
\bibitem [{\citenamefont {Schrinski}\ \emph {et~al.}(2017)\citenamefont
  {Schrinski}, \citenamefont {Stickler},\ and\ \citenamefont
  {Hornberger}}]{schrinski2017}%
  \BibitemOpen
  \bibfield  {author} {\bibinfo {author} {\bibfnamefont {B.}~\bibnamefont
  {Schrinski}}, \bibinfo {author} {\bibfnamefont {B.~A.}\ \bibnamefont
  {Stickler}}, \ and\ \bibinfo {author} {\bibfnamefont {K.}~\bibnamefont
  {Hornberger}},\ }\bibfield  {title} {\emph {\bibinfo {title}
  {Collapse-induced orientational localization of rigid rotors},\ }}\href@noop
  {} {\bibfield  {journal} {\bibinfo  {journal} {J. Opt. Soc. Am. B}\ }\textbf
  {\bibinfo {volume} {34}},\ \bibinfo {pages} {C1} (\bibinfo {year}
  {2017})}\BibitemShut {NoStop}%
\bibitem [{\citenamefont {Carlesso}\ \emph {et~al.}(2018)\citenamefont
  {Carlesso}, \citenamefont {Paternostro}, \citenamefont {Ulbricht},
  \citenamefont {Vinante},\ and\ \citenamefont {Bassi}}]{carlesso2018a}%
  \BibitemOpen
  \bibfield  {author} {\bibinfo {author} {\bibfnamefont {M.}~\bibnamefont
  {Carlesso}}, \bibinfo {author} {\bibfnamefont {M.}~\bibnamefont
  {Paternostro}}, \bibinfo {author} {\bibfnamefont {H.}~\bibnamefont
  {Ulbricht}}, \bibinfo {author} {\bibfnamefont {A.}~\bibnamefont {Vinante}}, \
  and\ \bibinfo {author} {\bibfnamefont {A.}~\bibnamefont {Bassi}},\ }\bibfield
   {title} {\emph {\bibinfo {title} {Non-interferometric test of the continuous
  spontaneous localization model based on rotational optomechanics},\
  }}\href@noop {} {\bibfield  {journal} {\bibinfo  {journal} {New J. Phys.}\
  }\textbf {\bibinfo {volume} {20}},\ \bibinfo {pages} {083022} (\bibinfo
  {year} {2018})}\BibitemShut {NoStop}%
\bibitem [{\citenamefont {Fadeev}\ \emph
  {et~al.}(2021{\natexlab{a}})\citenamefont {Fadeev}, \citenamefont
  {Timberlake}, \citenamefont {Wang}, \citenamefont {Vinante}, \citenamefont
  {Band}, \citenamefont {Budker}, \citenamefont {Sushkov}, \citenamefont
  {Ulbricht},\ and\ \citenamefont {Kimball}}]{fadeev2021ferromagnetic}%
  \BibitemOpen
  \bibfield  {author} {\bibinfo {author} {\bibfnamefont {P.}~\bibnamefont
  {Fadeev}}, \bibinfo {author} {\bibfnamefont {C.}~\bibnamefont {Timberlake}},
  \bibinfo {author} {\bibfnamefont {T.}~\bibnamefont {Wang}}, \bibinfo {author}
  {\bibfnamefont {A.}~\bibnamefont {Vinante}}, \bibinfo {author} {\bibfnamefont
  {Y.~B.}\ \bibnamefont {Band}}, \bibinfo {author} {\bibfnamefont
  {D.}~\bibnamefont {Budker}}, \bibinfo {author} {\bibfnamefont
  {A.}~\bibnamefont {Sushkov}}, \bibinfo {author} {\bibfnamefont
  {H.}~\bibnamefont {Ulbricht}}, \ and\ \bibinfo {author} {\bibfnamefont
  {D.~F.}\ \bibnamefont {Kimball}},\ }\bibfield  {title} {\emph {\bibinfo
  {title} {Ferromagnetic gyroscopes for tests of fundamental physics},\
  }}\href@noop {} {\bibfield  {journal} {\bibinfo  {journal} {Quant. Sci.
  Techn.}\ } (\bibinfo {year} {2021}{\natexlab{a}})}\BibitemShut {NoStop}%
\bibitem [{\citenamefont {Fadeev}\ \emph
  {et~al.}(2021{\natexlab{b}})\citenamefont {Fadeev}, \citenamefont {Wang},
  \citenamefont {Band}, \citenamefont {Budker}, \citenamefont {Graham},
  \citenamefont {Sushkov},\ and\ \citenamefont {Kimball}}]{fadeev2021gravity}%
  \BibitemOpen
  \bibfield  {author} {\bibinfo {author} {\bibfnamefont {P.}~\bibnamefont
  {Fadeev}}, \bibinfo {author} {\bibfnamefont {T.}~\bibnamefont {Wang}},
  \bibinfo {author} {\bibfnamefont {Y.~B.}\ \bibnamefont {Band}}, \bibinfo
  {author} {\bibfnamefont {D.}~\bibnamefont {Budker}}, \bibinfo {author}
  {\bibfnamefont {P.~W.}\ \bibnamefont {Graham}}, \bibinfo {author}
  {\bibfnamefont {A.~O.}\ \bibnamefont {Sushkov}}, \ and\ \bibinfo {author}
  {\bibfnamefont {D.~F.~J.}\ \bibnamefont {Kimball}},\ }\bibfield  {title}
  {\emph {\bibinfo {title} {Gravity probe spin: Prospects for measuring
  general-relativistic precession of intrinsic spin using a ferromagnetic
  gyroscope},\ }}\href {\doibase 10.1103/PhysRevD.103.044056} {\bibfield
  {journal} {\bibinfo  {journal} {Phys. Rev. D}\ }\textbf {\bibinfo {volume}
  {103}},\ \bibinfo {pages} {044056} (\bibinfo {year}
  {2021}{\natexlab{b}})}\BibitemShut {NoStop}%
\bibitem [{\citenamefont {Ma}\ \emph {et~al.}(2017)\citenamefont {Ma},
  \citenamefont {Hoang}, \citenamefont {Gong}, \citenamefont {Li},\ and\
  \citenamefont {Yin}}]{ma2016}%
  \BibitemOpen
  \bibfield  {author} {\bibinfo {author} {\bibfnamefont {Y.}~\bibnamefont
  {Ma}}, \bibinfo {author} {\bibfnamefont {T.~M.}\ \bibnamefont {Hoang}},
  \bibinfo {author} {\bibfnamefont {M.}~\bibnamefont {Gong}}, \bibinfo {author}
  {\bibfnamefont {T.}~\bibnamefont {Li}}, \ and\ \bibinfo {author}
  {\bibfnamefont {Z.-q.}\ \bibnamefont {Yin}},\ }\bibfield  {title} {\emph
  {\bibinfo {title} {Proposal for quantum many-body simulation and torsional
  matter-wave interferometry with a levitated nanodiamond},\ }}\href@noop {}
  {\bibfield  {journal} {\bibinfo  {journal} {Phys. Rev. A}\ }\textbf {\bibinfo
  {volume} {96}},\ \bibinfo {pages} {023827} (\bibinfo {year}
  {2017})}\BibitemShut {NoStop}%
\bibitem [{\citenamefont {Grimsmo}\ \emph {et~al.}(2020)\citenamefont
  {Grimsmo}, \citenamefont {Combes},\ and\ \citenamefont
  {Baragiola}}]{grimsmo2020}%
  \BibitemOpen
  \bibfield  {author} {\bibinfo {author} {\bibfnamefont {A.~L.}\ \bibnamefont
  {Grimsmo}}, \bibinfo {author} {\bibfnamefont {J.}~\bibnamefont {Combes}}, \
  and\ \bibinfo {author} {\bibfnamefont {B.~Q.}\ \bibnamefont {Baragiola}},\
  }\bibfield  {title} {\emph {\bibinfo {title} {Quantum computing with
  rotation-symmetric bosonic codes},\ }}\href {\doibase
  10.1103/PhysRevX.10.011058} {\bibfield  {journal} {\bibinfo  {journal} {Phys.
  Rev. X}\ }\textbf {\bibinfo {volume} {10}},\ \bibinfo {pages} {011058}
  (\bibinfo {year} {2020})}\BibitemShut {NoStop}%
\bibitem [{\citenamefont {Albert}\ \emph {et~al.}(2020)\citenamefont {Albert},
  \citenamefont {Covey},\ and\ \citenamefont {Preskill}}]{albert2020}%
  \BibitemOpen
  \bibfield  {author} {\bibinfo {author} {\bibfnamefont {V.~V.}\ \bibnamefont
  {Albert}}, \bibinfo {author} {\bibfnamefont {J.~P.}\ \bibnamefont {Covey}}, \
  and\ \bibinfo {author} {\bibfnamefont {J.}~\bibnamefont {Preskill}},\
  }\bibfield  {title} {\emph {\bibinfo {title} {Robust encoding of a qubit in a
  molecule},\ }}\href {\doibase 10.1103/PhysRevX.10.031050} {\bibfield
  {journal} {\bibinfo  {journal} {Phys. Rev. X}\ }\textbf {\bibinfo {volume}
  {10}},\ \bibinfo {pages} {031050} (\bibinfo {year} {2020})}\BibitemShut
  {NoStop}%
\end{thebibliography}
\end{document}